\documentclass{tADP2e}

\usepackage{epstopdf}
\usepackage{subfigure}

\theoremstyle{plain}

\theoremstyle{definition}

\theoremstyle{remark}

\begin{document}


\articletype{REVIEW}

\title{Formation of Exomoons: A Solar System Perspective}

\author{
\name{Amy C. Barr$^{\ast}$\thanks{$^\ast$Corresponding author. Email: amy@psi.edu}}
\affil{Planetary Science Institute, 1700 East Ft. Lowell Rd., Suite 106, Tucson AZ 85719 USA}
\received{Submitted Sept 15, 2016, Revised October 31, 2016}
}

\maketitle

\begin{abstract}
Satellite formation is a natural by-product of planet formation.  With the discovery of numerous extrasolar planets, it is likely that moons of extrasolar planets (exomoons) will soon be discovered.  Some of the most promising techniques can yield both the mass and radius of the moon.  Here, I review recent ideas about the formation of moons in our Solar System, and discuss the prospects of extrapolating these theories to predict the sizes of moons that may be discovered around extrasolar planets.  It seems likely that planet-planet collisions could create satellites around rocky or icy planets which are large enough to be detected by currently available techniques.  Detectable exomoons around gas giants may be able to form by co-accretion or capture, but determining the upper limit on likely moon masses at gas giant planets requires more detailed, modern simulations of these processes.
\end{abstract}



\section{Introduction}
The discovery of a bounty of extrasolar planets has raised the question of whether any of these planets might harbor moons.  The mass and radius of a moon (or moons) of an extrasolar planet (exomoon) and its host planet can offer a unique window into the timing, duration, and dynamical environment of planet formation, just as the moons in our Solar System have yielded clues about the formation of our planets \cite{Agnor1999,Chambers2004,cr,Barr:2010ab,Sasaki2010}.  Exomoons may also provide conditions conducive to the presence of liquid water in planetary systems where it might not otherwise be present \cite{Reynolds1987, Williams1997, Kaltenegger2010, Heller2012, HellerBarnes2013}, expanding the number of possible abodes for life in the galaxy \cite{Tarter2007}.

In our Solar System, planetary satellite systems fall into two broad classes when analyzed in terms of their total satellite system-to-planet mass ratio ($M_{S}/M_{pl}$).  For rocky and icy planets (those without substantial gaseous envelopes), $M_S/M_{pl} \sim 0.01$ to 0.1.  These bounds are set by Earth's Moon and Charon, the large satellite of the dwarf planet Pluto.  The small moons of Mars, Phobos and Deimos, which may be only short-lived \cite{Burns1978}, have $M_S/M_{pl}\sim10^{-8}$.  For the outer planets, each of which has a substantial gaseous envelope, $M_S/M_{pl} \sim 10^{-4}$.  

In the context of modern planet formation theory, the formation of satellites is a natural by-product of the formation of planets, both terrestrial and gas-rich.  Satellites of the terrestrial planets and outer planets are thought to have formed by different mechanisms.  Earth's Moon and Charon are thought to have formed from planet-scale collisions \cite{HartmannDavis, CameronWard, CanupAsphaug, Canup2004, Canup2005}.  Numerical simulations of the collisions and the subsequent accretion of impact-generated debris show that systems with the proper masses, angular momenta, and compositions of each system are consistent with a collision with a body 10\% to 50\% the mass of the precursor planet \cite{Canup2004, Canup2005, Canup2011, StewartCuk2012, Canup2012}.  In the inner Solar System, collisions between planetary embryos are common during the late stages of terrestrial planet accretion \cite{Agnor1999, Chambers2004, Chambers2013}.  It is logical to assume that terrestrial planets formed in other systems would experience such collisions as well \cite{OgiharaIda2009, Morishima2010, Elser2011}.  

The satellites of the outer planets are thought to form by co-accretion \cite{PollackConsolmagnoSaturnBook, Coradini1989, CW02, ME03}, in a circumplanetary disk of material pulled into orbit while the young planet is accreting gas from the solar nebula \cite{CW02, WardCanup2010}.  In one of the more successful formulations of the co-accretion theory, the satellite system-to-total mass ratio shared by all four planet/satellite systems is thought to arise from a balance between satellite loss by Type I migration and satellite growth from sweep-up of solids  in the mid-plane of the circumplanetary disk \cite{CW02, CW06}.  The value of $M_S/M_{pl}$ is relatively insensitive to planetary mass \cite{CW06}, suggesting that primordial satellite systems around extrasolar planets could have $M_S/M_{pl}\sim10^{-4}$ as well \cite{Heller2015a, Heller2015b}.  

There are several methods by which exomoons may be detected, including transits \cite{Kipping2012-1, Heller2014Detection, Hippke}, pulsar time of arrival analysis \cite{LewisPulsar}, modulation of planetary radio emissions \cite{Noyola},  gravitational microlensing \cite{Liebig}, time-resolved imaging polarimetry \cite{SenguptaMarley}, and indirect means \cite{Mamajek2015}.  Of these, transit detection methods are among the most promising, given the abundant observations from \emph{Kepler} and plans for future space-based observatories, and because the detection of the effect of an exomoon on the timing and duration of transits can yield both a mass and radius for the moon \cite{Kipping2009b}.  Coupled with the mass and radius of the host planet, one can begin to deduce basic information about how the moon may have formed.  The effect of an exomoon (or moons) on the behavior of a transiting planet is strongly dependent on the satellite-to-planet mass ratio \cite{Sartoretti1999, Szabo2006, Kipping2009a, Kipping2009b}, so systems with a single large satellite are the most likely to be found.  

The Hunt for Exomoons with \emph{Kepler} (HEK) team reports that their survey can plausibly detect a planetary moon $\gtrsim 0.1$ Earth masses, or roughly the mass of Mars.  This is about ten times larger than the largest planetary satellites in our Solar System.  At present, it is not clear how moons of this size may form \cite{Kipping2012-1}, or how common they might be.  This is because very little is known about how the satellite formation processes proposed for our Solar System might scale to different planetary masses and stellar conditions.  

Here, I discuss the present state of knowledge of satellite formation in our Solar System, and prospects for extrapolating these theories to predict satellite sizes in exoplanetary systems.  In Section \ref{sec:SS}, I discuss the classes of satellites observed in our Solar System.  Formation of satellites around rocky and icy planets is described in Section \ref{sec:impacts}.  Gas giant satellite formation is discussed in Section \ref{sec:co-accretion}.  Satellite formation methods proposed for planetary satellites which have been ruled out because they do not match one or more Solar System observations are discussed in Section \ref{sec:other-means}.  Exomoon detection methods are summarized in Section \ref{sec:detection}.  The paper concludes with a discussion about the possibilities for, and limitations of extrapolating Solar System theories to other systems, and avenues for future work.

\section{Satellites in our Solar System \label{sec:SS}}
Table \ref{table:satellites} lists the largest satellites of each of the planets in our Solar System, including the satellites of the dwarf planet Pluto.  Any object executing a prograde orbit that is roughly co-planar with the equator of a larger planetary body is considered to be a regular satellite.  Thus, Earth's Moon, Phobos and Deimos at Mars, the satellites of Pluto, and the large moons of the outer planets are considered to be regular satellites.  The semi-major axes of the major regular satellites range from a few to a few tens of planetary radii, indicating that the satellites are deeply embedded in their planet's Hill spheres.  This means that the gravitational force from the planet is overwhelmingly dominant over gravity from the central star.  This also indicates that the satellites are tightly bound to the planet; satellite systems orbiting this close to the planet could likely survive planetary migration \cite{Domingos2006, Namouni2010}, unless the planet migrates very close to the central star \cite{BarnesOBrien2002}. 

Irregular satellites are those on inclined, eccentric, and/or retrograde orbits.  Often, irregular satellites are $\sim 100$ km or smaller objects that have been captured in the gravitational well of the host planet \cite{Nesvorny2007}.  Neptune's large retrograde satellite Triton, thought to be a captured object, is an oddity that will be discussed separately, in Section \ref{sec:oddities} below. Each of the outer planets has several families of irregular satellites.  Typically there is one family of small objects that orbit between 2 and 6 planetary radii from the parent body and are on low-inclination orbits, and one or more additional families that orbit a hundred to a thousand planetary radii from the host planet on highly inclined orbits \cite{StevensonSatsBook}.  Although some of the small irregular satellites and Trojan asteroids in our Solar System may be close in size to the regular satellites of the outer planets, detecting them would be difficult using current instrumentation, unless they exist in enormous numbers.  This could occur, for example, in a system that has been recently disrupted by a recent stellar encounter, causing the planets within to collide and create tremendous numbers of small fragments \citep{BodmanKIC2016}.

Table \ref{table:satellites} also lists properties of the satellite systems of our Solar System that provide clues about how the systems formed.  These include the orbital radius of each satellite, scaled by planetary radius, the character of the orbit (e.g., whether the orbit is regular and the satellite is tidally evolved \cite{StevensonSatsBook} or the satellite is on an inclined orbit), and the ratio of the density of the satellite to the density of the host planet.  

\subsection{Orbits}
A prograde orbit with low inclination is a strong indicator that a planetary satellite formed from material orbiting the host planet.  The orbits of some satellites are also tidally evolved, meaning that the tidal torques between the planet and satellite have acted to equate the orbital period of the satellite with its spin period.  This dynamical state, called the ``synchronous'' state, is achieved in the first few hundred million years after formation (e.g., \cite{DobroPlutoBook}).  In this state, the satellite shows the same hemisphere to the planet.   Earth's Moon is currently in a synchronous state.  Over longer time scales, tidal torques transfer the spin angular momentum of the parent planet to the orbital angular momentum of the satellites, evolving the system toward a ``dual synchronous'' state in which the spin period of the planet is equal to the spin and orbital period of the satellite.  The only place in our Solar System where this has occurred is the Pluto system.  

\subsection{Mass and Composition}
The ratio between the satellite and planets' mean densities and masses can also be diagnostic of the mode of satellite origin.  The satellite systems of the outer planets have $M_{S}/M_{pl}\sim10^{-4}$ \cite{CW06}.  Earth's Moon and Pluto's Charon are larger relative to their primaries: the Moon is about 1\% of the mass of the Earth, and Charon is about 12\% the mass of Pluto.  

\begin{figure}
\includegraphics[width=160mm]{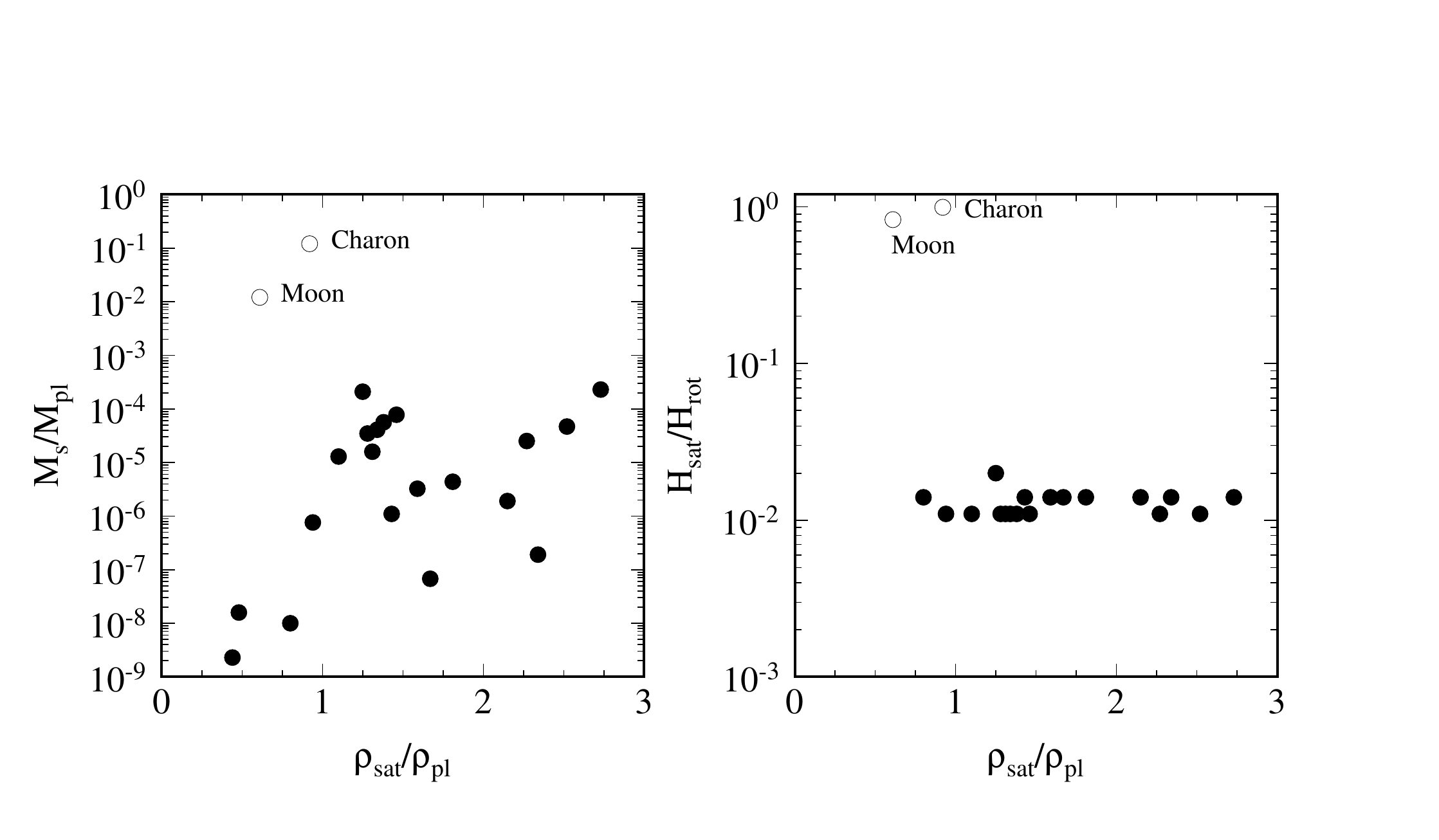}
\caption{Satellites of our Solar System's planets (dots), as a function of satellite-to-planet mass ratio (left), and ratio of total angular momentum in the orbital motion of the satellites, $H_{sat}$ to the total system angular momentum, $H_{sat}+H_{rot}$ \cite{MacDonald1966}.  Satellites thought to have formed from a giant impact, Earth's Moon, and Pluto's satellite Charon, are shown as open circles. \label{fig:density}}
\end{figure}

The outer planets are gas-rich and have low mean densities, $\sim 0.6$ to $1.6$ g/cm$^{3}$, but their satellites are composed of mixtures of ice and rock, with densities $\sim 0.8$ to $1.9$ g/cm$^3$. Thus, $\rho_{sats}/\rho_{pl} >1$.  The satellites of rocky/icy planets (Earth, Mars, and Pluto) are comparable in density or less dense than their host planets ($\rho_{sats}/\rho_{pl} < 1$).  The left panel of Figure \ref{fig:density} plots the satellites of our Solar System in terms of $\rho_{sats}/\rho_{pl}$, and the ratio between satellite mass and planet mass, $M_{s}/M_{pl}$.  

\subsection{Angular Momentum}
Table \ref{table:satellites} also lists the ratio between the angular momentum contained in the orbital motion of the satellites, $H_{sat}$, and the total system angular momentum, $H_{tot}=H_{sat}+H_{rot}$, where $H_{rot}$ is the rotational angular momentum of the planet.  The angular momentum for each of the satellites' orbits, $H_{sat}=\mu \sqrt{a G(M_{s}+M_{pl})}$, where $G$ is the gravitational constant, $a$ is the semi-major axis of the satellite's orbit, and $\mu=M_sM_{pl}/(M_s+M_{pl})$ is the reduced mass.  For the host planet, $H_{rot}=I\omega_{pl}$, where $I=\alpha M_{pl}R_{pl}^2$ is the planet's moment of inertia, $\alpha$ is its polar moment of inertia coefficient, and $\omega_{pl}$ is its spin frequency.  The right panel of Figure \ref{fig:density} illustrates how the angular momentum ratio for our Solar System's satellites vary as a function of density ratio.  Again, the Moon and Charon stand out as anomalous, with $H_{sat}/H_{tot}\sim1$, and $H_{sat}/H_{tot}\sim 0.01$ for the satellites of the gas-rich outer planets.

\subsection{Oddities \label{sec:oddities}}
In terms of their basic properties, two satellite systems stand out as odd: Phobos and Deimos at Mars, and Triton, the large satellite of Neptune that is in an inclined, retrograde orbit.  Phobos and Deimos have $M_{s}/M_{pl}\sim 10^{-8}$ to $10^{-9}$, and densities lower than the bulk density of Mars.  Triton orbits in the opposite sense to Neptune's rotation, at an inclination of 157$^{\circ}$ relative to Neptune's equator.  This points toward Triton being a captured body \cite{McKinnon1984, GoldreichNeptune, AgnorHamiltonTriton}.  

Although one might be tempted to dismiss these cases as ``odd'' in the context of satellite formation in our Solar System, it is important to note that the first satellites we find are likely to be ``oddities'' because large moons unlike those in our Solar System are the only types of systems we can observe with present techniques (e.g., \citep{Kipping2012-1}).  For example, the first extrasolar planets discovered were objects with several earth masses found orbiting pulsars \citep{WolszczanFrail}, whose existence would not have been predicted by standard Solar System evolution models.


\section{Satellite Formation via Giant Impact \label{sec:impacts}}
In our Solar System, embryo impacts onto the young inner planets are thought to play a crucial role in setting their final chemical compositions and dynamical states.  Impacts between planetary embryos of 0.1 to 1 Earth-mass were thought to be common during the late stages of planet formation in our Solar System, 200-300 Myr after the dissipation of the solar nebula (see e.g., \cite{Morby2012} for discussion).   A collision or multiple collisions have been suggested as an explanation for the anomalously iron-rich composition of the planet Mercury \cite{AsphaugReufer2014}, the slow rotation rate and \emph{absence} of moon for Venus \cite{AlemiStevenson}, Earth's large moon \cite{HartmannDavis, CameronWard, CanupAsphaug, Canup2004}, and the crustal dichotomy \cite{Marinova2008} and small satellites of Mars \cite{BurnsMars, Peale2007, Craddock2011, CitronMarsMoons}.  Pluto's satellites probably formed as a result of an impact between two Kuiper Belt objects \cite{Canup2005, Canup2011, KenyonBromley2013}.  It also seems likely that a phase of embryo-embryo collisions would have occurred in other solar systems, as well (e.g., \cite{OgiharaIda2009, Morishima2010}). 

Determining whether a planetary collision will form a satellite (or satellites), and the sizes of the resulting object(s) requires simulating two processes: (1) the collision itself, and (2) the evolution of the impact-generated debris.  The mass of debris placed into orbit during the impact can be viewed as an upper limit on the mass of satellite(s) \citep{Elser2011}.  But some of the orbiting material will re-impact the planet or may be lost.  

\subsection{Collisions between Rocky Planets}
\subsubsection{Impact Simulations}
To date, most numerical simulations of impacts between rock/metal protoplanets have been performed with the goal of reproducing the mass, angular momentum, and composition of the Earth-Moon system (e.g., \cite{BenzMoon, CanupAsphaug, Canup2004, Elser2011, Canup2012, StewartCuk2012, ReuferMoon2012, Meier2014, Hyodo}). Early simulations of the Moon-forming impact were forced, due to computational constraints, to use low numerical resolution, or to use simplified equations of state, leading to unphysical behaviors for rock and metal at high temperatures and pressures.  For this reason, numerical simulations of terrestrial planet impacts are still focused on determining the conditions under which the Moon formed (see Barr (2016) \citep{moon_review} for discussion).

Impact velocities for embryo-embryo collisions are comparable to the mutual escape velocity of the colliding objects, $v_{esc,sys} = \sqrt{2 G(M_i+M_t)/(R_i+R_t)}$, where $M_i$ and $R_i$ are the mass and radius of the smaller body (the ``impactor'') and $M_t$ and $R_t$ are the mass and radius of the larger body (the ``target'').  Impacts occurring at a few km/s, such as those between icy or rocky bodies with radii greater than a few hundred kilometers, will launch a shock wave in the target and impactor, and generate significant vapor and melting.  To obtain an accurate estimate of the mass of orbiting material, impacts of this type need to be simulated using shock physics codes such as the smoothed particle hydrodynamics (SPH; a Lagrangian method) \cite{Monaghan1992, BenzMoonPaper1}, Eulerian methods \cite{WadaNorman2001, Wada2006}, or hybrid Eulerian/Lagrangian methods such as the hydrocode CTH, commercially available from Sandia National Laboratories \cite{McGlaun, Crawford2006, m-moon}.  

\begin{figure}
\includegraphics[width=140mm]{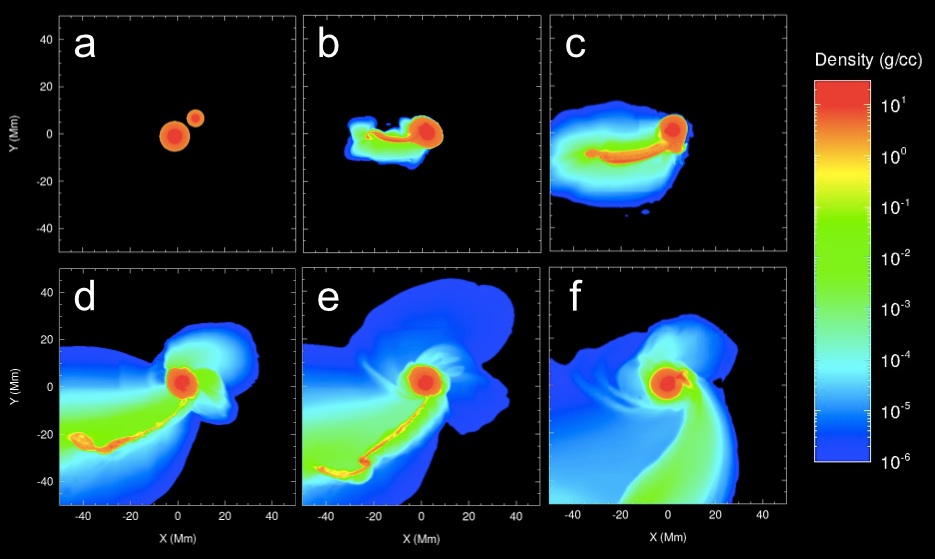}
\caption{CTH simulation of the Moon-forming impact, after Canup et al., (2013) \cite{m-moon}; details about the simulation method, equations of state, and data analysis may be found in this reference.  A Mars-sized object composed of 70\% rock and 30\% iron collides with the protoearth of similar bulk composition, with a $46^{\circ}$ impact angle, at $v_{imp}=v_{esc,sys}$.  Colors indicate density in g/cm$^3$, on a logarithmic scale.  Simulation results are plotted as a function of time for $t=0$ hours (a), 1 hour (b), 1.8 hours (c), 3.4 hours (d), 4.8 hours (e), and 10 hours (f) after the impact.  The impact launches a disk of $M_{o}\sim 1.2$ lunar masses worth of silicate-rich material into orbit.\label{fig:ser119}}
\end{figure}

Figure \ref{fig:ser119} illustrates, as an example of a collision between two rocky planets, the first 10 hours of a CTH simulation of the Moon-forming impact \cite{m-moon}.  Details about the numerical method, inputs, and equations of state may be found in Canup et al., (2013) \cite{m-moon}.  The total colliding mass $M_T\sim1.02M_E$.  The impactor-to-total mass ratio, $\gamma=M_i/M_T$ is equal to 0.13.  The objects collide with an angle of 46$^{\circ}$, at an impact velocity $v_{imp}\sim v_{esc,sys}$.  In the early stages of the impact, the impactor is sheared out into an arm of material which collapses back onto the protoearth about 10 hours after the initial impact.  The iron core of the impactor and protoearth merge, and a large orbiting disk of heated silicate-rich material is left orbiting the protoearth, which has been extensively internally perturbed and heated.

The subsequent evolution of the debris into a final moon or moons can be simulated using $N$-body techniques \cite{IdaCanupStewart, Kokubo2000, Hyodo}.  Collisions between large protoplanets generate extremely hot disks which may contain a mixture of silicate vapor and melt \cite{ThompsonStevenson}.  Determining the mass of satellite formed from such a disk requires simulating self-consistently modeling the cooling and spreading of the disk close to the planet, and subsequent gravitational sweep-up of debris orbiting outside the planet's Roche limit, far enough from the planet that its gravity does not prevent accretion \cite{Ward2011, SalmonCanup, SalmonCanup2014}.  

Collisions designed to re-create the observed properties of the Earth/Moon system typically involve $M_T \sim 1.05 M_E$, and impact velocities comparable to the escape velocity of the system, $v_{imp} \sim v_{esc,sys} \sim 10$ km/s \cite{Canup2004, Canup2008}.  Large impact-generated disks can form in collisions with higher velocities, a few$\times v_{esc}$, but this phenomenon has been explored for only a limited range of impactor-to-target mass ratios, and impact angles \cite{StewartCuk2012, Canup2012}.  

To date, very few hydrocode simulations of satellite formation via giant impact have been performed for other systems, and none have been performed for impacts involving a total mass $M_T>>M_E$.  Thus, not much is known about how the process scales with planetary mass, or with impact velocity, angle, or impactor-to-target mass ratio \cite{Elser2011}. 

\subsubsection{Disk Mass}
The mass of a moon formed in a collision depends on the amount of material launched into orbit during an impact, $M_o$, its angular momentum, $L_o$, and the efficiency with which that material is accreted into a final moon (or moons) \cite{IdaCanupStewart}.  

The final $M_o$ depends on the mass of the lens-shaped region of contact representing the overlap between the spherical target and impactor, $M_{interact}$ \cite{Canup2008}, 
\begin{equation}
\frac{M_o}{(M_t+M_i)} \sim C_{\gamma} \bigg(\frac{M_i - M_{interact}}{(M_t+M_i)}\bigg)^2, \label{eq:diskmass}
\end{equation}
with the factor $C_{\gamma} \sim 2.8 (0.1/\gamma)^{1.25}$ determined empirically based on the results of impact simulations, and the ratio between the impactor mass and total mass involved in the collision, $\gamma=M_i/(M_t+M_i)$.  Thus, if one knew $C_{\gamma}$, it would be possible to predict $M_o$ without having to simulate the impact.  Unfortunately, this scaling relationship (in particular the value of $C_{\gamma}$) has only been determined for a narrow range of impact conditions, $0.1 \lesssim \gamma \lesssim 0.2$, moderate impact angles, and $1 \lesssim v_{imp}/v_{esc} \lesssim 1.4$.  Its applicability to impacts beyond this range is unclear \cite{Elser2011}.  

Some numerical simulations seeking to relate impactor properties to outcomes have been performed for conditions very much unlike the Moon-forming impact, but their results are not directly applicable to the formation of satellites.  Numerical simulations for extremely low impact velocities relevant to the accretion of kilometer-sized planetesimals have provided additional information about how the mass of the largest remnant in a collision scales with $M_{interact}$ for a wide range of impact conditions, but does not report the mass of bound, orbiting material \cite{Leinhardt2012I}.  Additionally, the Leinhardt et al., \cite{Leinhardt2012I} simulations do not include shock propagation, melting, or vaporization, all of which will affect energy partitioning and ejection \cite{GH, StevensonSatsBook}.  Numerical simulations for super- and sub-Earth mass collisions with $v_{esc} < v_{imp} < 6v_{esc}$ also highlight the importance of the size of the high-density planetary core in determining the mass of orbiting material and its composition \cite{MarcusRocky2009, MarcusIcy2010, Genda2012}.  It seems likely that $C_{\gamma}$ will change as the sizes of the cores of the target and projectile change.

\subsubsection{Accretion from a Particulate Disk\label{sec:gas-free}}
\begin{figure}
\includegraphics[width=80mm]{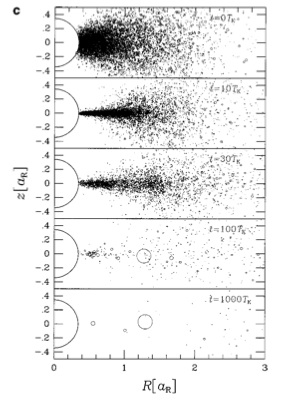}
\caption{Reproduced with permission from Kokubo et al. (2000) \cite{Kokubo2000}.  Time-evolution of a gas-free disk of debris after the Moon-forming impact (simulation 15a).  Four lunar masses of material is initially distributed in a region 2 Roche radii from the Earth (about 6 Earth radii).  The particles obey a size distribution $n\varpropto m^{-1.5}$, where $n$ is the number of particles, and $m$ is particle mass.  If the particles all of equal mass, they would have a radius $r\sim 130$ km, assuming a density of 3.3 g/cm$^3$ for the disk material. Time is reported in units of the orbital period at the Roche limit, $T_K=7$ hours for the Earth.  Over the entire duration of the simulation, 1000$T_K$ (or 290 days, where 1 day is 24 hours), a single large Moon accretes just outside the Roche limit.  \label{fig:Kokubo2000}}
\end{figure}

The mass of material orbiting the planet after the impact gives an \emph{upper limit} on the mass of satellite or satellites that can form \cite{Elser2011}.  However, determining the actual mass of the satellite, and how many satellites will form, requires simulating the sweep-up of debris. The simplest way of simulating accretion is to neglect the presence of gas or vapor, and assume that all of the debris is solid (a ``particulate disk.'')  In this case, the evolution of the debris is governed solely by gravity \cite{IdaCanupStewart, Kokubo2000, TakedaIda, Hyodo}.  Figure \ref{fig:Kokubo2000} illustrates the accretion of the Moon in a gas-free disk originally containing 4 lunar masses of material distributed within 2 Roche radii of the Earth.  The simulations show that the impact-generated debris accretes into a large satellite very quickly, in less than a year.  

Simulations of this type have been performed for a wide range of $M_o$ values.  In particulate disks with $M_o/M_{p} > 0.03$, common in the Moon-forming impact, accretion of a single satellite is favored \cite{IdaCanupStewart, Kokubo2000, Hyodo}, and its mass ($M_L$) is proportional to the ratio of disk angular momentum to the angular momentum of an orbit at the Roche limit (radius $a_R$) \cite{Kokubo2000},
\begin{equation}
\frac{M_L}{M_o} \approx a \bigg(\frac{L_o}{M_o \sqrt{GM_{p} a_R}}\bigg) - b - c \bigg(\frac{M_{esc}}{M_o}\bigg), \label{eq:ICS97}
\end{equation}
where $L_o$ is the angular momentum of orbiting material, and the amount of mass that escapes the system during accretion, $M_{esc} \leq 5$\%.  The numerical coefficients $a=1.9$, $b=1.15$, and $c=1.9$ are obtained from fitting the outcomes of many $N$-body simulations of the sweep-up a purely solid (particulate) impact-generated disk.  In a typical simulation of the Moon-forming impact, very little mass is lost, so that $M_{esc}/M_o < 0.05$ \cite{Kokubo2000}.  

\begin{figure}
\includegraphics[width=140mm]{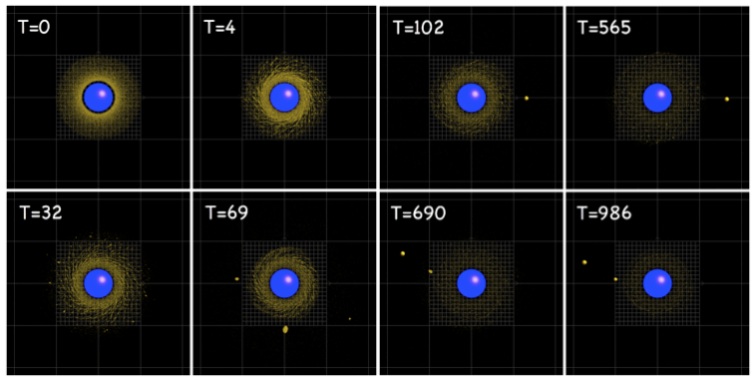}
\caption{Reproduced, with permission from Figure 5 of \cite{Hyodo}.  Snapshots of the gas-free accretion of solid debris after an impact between two rock/metal planets.  The disk initially contains 1.25 lunar masses of material, with specific angular momentum $j=0.775$.  Results are shown looking down onto the disk from above.  The disk material is initially distributed between 0.4 and 1 Roche radius from the planet, which has a radius 0.34 Roche radii ($\sim 1$ Earth radius).  Numbers indicate elapsed time in units of the orbital period at the Roche limit, which for Earth, is about 7 hours. \label{fig:hyodo}}
\end{figure}

Simulations of accretion in higher-mass disks, with $0.003 < M_o/M_{p} < 0.03$, and a narrow range of $L_o$, show that formation of two satellites can be favored over a single large satellite \cite{Hyodo}, especially in disks with low angular momentum.  Figure \ref{fig:hyodo} illustrates the formation of two moons in a disk containing 1.25 lunar masses of material initially distributed inside the Roche limit of an Earth-mass planet.  The disk contains a specific angular momentum $j_{disk}=0.775$, where $j_{disk}=L_o/L'$, and $L'=\sqrt{GM_{pl}R_{pl}}$ \cite{Hyodo}.  These conditions are similar to those expected after the Moon-forming impact.  Over a period of a little less than a year, the material accretes into two moons which orbit the Earth in a 2:1 mean motion resonance \cite{Hyodo}.  The outer satellite has 15\% of a lunar mass, and the inner satellite has about 10\% of a lunar mass.  The mass of the first satellite is proportional to $M_o^2$.  The two satellites can be comparable in mass, or quite dissimilar in size, and can form in a 2:1 mean motion resonance, or be co-orbital \cite{Kokubo2000, Hyodo}.  

For $M_o/M_{p} < 0.003$, $N$-body simulations of disks with low mass but constant surface-mass density (not a realistic assumption for impact-generated disks) show that $M_L/M_{p} \sim 2200(M_o/M_{pl})^3$ \cite{CridaCharnoz}.  Although these analyses provide a general framework for relating disk mass to satellite masses, more $N$-body simulations are required to explore a larger range of $M_o$ and $L_o$.

\subsubsection{Accretion from a Liquid/Vapor Disk}
Although a particulate disk model can yield insight into the mass of the Moon, it does not provide realistic constraints on the time scale of lunar accretion or the geochemical evolution of the disk.  A collision between two sizable rock/iron planetary bodies creates an orbiting disk of material that is composed of a two-phase silicate ``foam'' Ð containing both liquid and vapor \cite{WardCameron1978, ThompsonStevenson, Canup2004}.  The mass of the final moon (or moons) created, and the time scale over which accretion takes place, is controlled by the evolution of the disk vapor \cite{ThompsonStevenson, Ward2011, SalmonCanup, SalmonCanup2014}.  

The impact-generated disk extends from the surface of the Earth itself to a distance of $\sim$ a few to ten Earth radii.  Inside the Roche limit, $\sim 3$ Earth radii, gravitational tidal forces from the planet are too high to permit accretion.  As the disk cools, it condenses.  However, strong tidal dissipation in the dense disk can cause material to vaporize.  The balance between dissipation and energy loss by radiation controls the time scale over which the disk spreads \cite{Ward2011}. Outside the Roche limit, material initially placed into orbit by the impact is quickly swept up into a single object, a ``nucleus'' of the Moon, which has about 40\% of a lunar mass \cite{SalmonCanup}.  Over the next 100-1000 years, while the Roche-interior disk cools and spreads, small fragments spawned off the edge of the Roche zone accrete onto the original nucleus of the Moon.  

Coupled simulations of the evolution of the liquid/vapor disk and $N$-body simulations of the sweep-up of fragments outside the Roche zone reveal that the moons created from a two-phase disk are slightly smaller than those obtained in simulations assuming a purely particulate disk \cite{SalmonCanup}.  The results of these more sophisticated simulations yield different values for the coefficients in equation (\ref{eq:ICS97}), $a=1.14$, $b=0.67$, and $c=2.3$ \cite{SalmonCanup}.  

Of course, these simulations have only been performed for conditions similar to the Earth-Moon impact.  To truly understand the mass and chemistry of a resultant moon, these techniques should be applied to the study of more massive disks around more massive planets, at a variety of temperature conditions, and for a variety of compositions, including water- and/or metal-rich disks.

\subsection{Pluto/Charon \label{sec:Pluto}}
Like Earth's Moon, Pluto's satellite Charon, is also anomalously large relative to Pluto, with $M_s/M_{pl}=0.12$.  Charon also contains the vast majority of the system's angular momentum, with 99\% being presently contained in Charon's orbit.  Charon is also in a dual-synchronous state, in which its spin period, orbital period, and Pluto's spin period are all identical.  All of these properties suggest that Charon may have formed during a collision between two like-sized Kuiper Belt objects \cite{Canup2005, Canup2011}.  

Smoothed particle hydrodynamics simulations of impacts between Kuiper Belt objects show that systems with the proper mass ratio and angular momentum can be created in an impact with $\gamma=0.3$, $v_{imp}=v_{esc}$, and an impact angle of 73$^{\circ}$.  Interestingly, the interior state of the Pluto/Charon precursor bodies has a significant effect on Charon's composition.  In collisions between two fully differentiated objects (i.e., objects with a rock or rock/metal core and ice mantle), the cores merge, causing the moment of inertia of the merged central body to decrease.  The central body begins to spin rapidly, lofting mantle ice off the surface \cite{Canup2005, Canup2011}.  The result is a disk of ice-rich material, which is inconsistent with Charon's relatively rock-rich composition.  Thus, a grazing impact with an undifferentiated ice/rock impactor is favored; in this case, Charon represents a largely intact fragment of the impactor \cite{Canup2011}.

As the \emph{New Horizons} spacecraft approached Pluto, new small satellites were discovered \cite{Weaver2006, Showalter2011}, raising the question of whether an impact could create an intact Charon with the proper composition and also a disk of material that could form the small satellites.  Subsequent impact simulations show that a slightly faster collision, with impact angles greater than $\sim 48^{\circ}$ can yield both a disk and an intact satellite.  The sweep-up and dynamical evolution of the disk can explain the masses, orbits, and seemingly icy compositions of the small moons \citep{KenyonBromley2013}.

It is worth noting that several large Kuiper Belt Objects have ice-rich satellites that are much smaller than Charon.  These Kuiper Belt objects also have rock-rich compositions, suggesting that they may have suffered one or more ice-stripping collisions \citep{LeinhardtHaumea, densities_paper}.  More work on this type of impact could shed light on the origin of these rock-rich dwarf planets, but also open the possibility of forming iron-rich exoplanets with rock-rich moons.
 
\section{Co-Accretion \label{sec:co-accretion}}

\subsection{Minimum-Mass Subnebula}
The major satellites of Jupiter, Saturn, and Uranus are in prograde, low-inclination orbits, deeply embedded in the planets' Hill spheres, suggesting that they formed in an orbiting disk of material.  Several models for the origin and evolution of the disk have been proposed. The classical model uses the masses of the satellites themselves to construct a ``minimum mass sub-nebula'' (MMSN), analogous to a scaled-down version of the minimum mass solar nebula (e.g., \cite{PollackConsolmagnoSaturnBook}). In the MMSN model, the surface mass density of gas and solids as a function of distance away from the planet is constructed based on the satellites' present masses and locations. 

The MMSN model is a logical starting point for understanding the structure of a circumplanetary disk, but it has several serious flaws.  A jovian MMSN has a monotonic decay of surface mass density with distance, but MMSN models constructed from the masses and locations of the saturnian and uranian satellites do not (see Figure 1 of \cite{CW02}). Densities in a jovian or saturninan MMSN can reach as high as $\sigma \sim 10^3$ to $10^5$ g/cm$^3$ \cite{Squyres88, Coradini1989}, several orders of magnitude higher than the solar nebula.  In such a dense disk, which is heated by thermal emission from the contracting giant planet at its center, it is difficult to obtain temperatures low enough to accrete ice-rich moons \cite{CW02}.  Modern and more realistic treatments of the MMSN implement physically realistic power-law decay curves for surface mass density and temperature \cite{ME03}, but find that MMSN disks are optically thick, which still give temperatures too high to accrete ice-rich satellites \cite{CW02} unless the viscosity of the disk is quite small \cite{ME03}.  Moreover, creating the MMSN structure within a self-consistent model of gas flow and contraction of a giant planet is impossible \cite{Ward2010}.  

\subsection{The Gas-Starved Disk}
In the last decade, new simulations of gas giant formation have motivated an alternative model for the circumplanetary disk.  Simulations of gas accretion onto growing giant planet cores show that the planets with masses less than 30 Jupiter masses, with and without deuterium burning, contract to approximately their final size after about a million years of growth \cite{PapaloizouNelson2005, Hubickyj2005, MolliereMordasini2012, Mordasini2013}, much shorter than the nominal lifetime of disks around stars $\sim 10$ Myr \citep{Haisch2001, Thi}. For much of the planet's growth, gas flowing into the Hill sphere that has too much angular momentum to accrete will achieve planetary orbit and form a circumplanetary disk \cite{Lubow99, DAngelo2003, AyliffeBatePaper2, TanigawaPaper1} (see Figure \ref{fig:Tanigawa} here).  

The gas also contains $\sim$meter-sized particles of rock and ice that are hydrodynamically coupled to the gas and also enters orbit in the disk \cite{TanigawaPaper2}. The gas flows viscously onto the planet and outward where it can be lost from planetary orbit \cite{CW02}. The solids quickly accrete into bodies large enough to become de-coupled from the gas, and so they remain in orbit where they are quickly swept up into satellites \cite{CW02,CW06} (see also Figure \ref{fig:gsd} here).  In such a disk, the instantaneous surface mass density in the disk is low, representing a steady state between gas added from the nebula and lost from the edges of the disk and to the planet; for this reason, the model is referred to as a ``gas-starved'' disk.  

The gas-starved disk is an attractive model to pursue when looking at the possibility of exomoon formation because it is the only co-accretion model that is predicative.  The MMSN and variants of it require knowledge of the sizes and orbital spacings of the satellites to construct the surface mass density profile of the disk.  The gas-starved disk predicts a maximum satellite mass and total mass of satellite system, which is largely insensitive to disk conditions for jovian and saturnian conditions. Solid particles in the disk are subject to gas drag, Type I orbital migration, and eccentricity and inclination damping \cite{PapaloizouLarwood, TanakaWard2004, CresswellNelson}. The maximum satellite mass is dictated by the balance between the timescale of inward Type I migration and the accretion time scale of a growing satellite, which gives $M_{lg} \sim 5.6 \times 10^{-5}M_{pl}$, and a total mass of the satellite system $M_T \sim 2.5\times10^{-4} M_{pl}$ \cite{CW06}. As the gas inflow from the solar nebula wanes, inward migration slows and eventually stops, leaving behind the system's final satellites. If gas inflow decreases slowly over time due to the gradual dissipation of the solar nebula, the jovian- and saturnian-mass systems tend to form a single $M_{lg}$ body and a half-dozen smaller satellites; if gas inflow ceases abruptly because the planet opens a gap in the solar nebula, the system tends to form three to seven bodies with $M_s \sim M_{lg}$ \cite{Sasaki2010}. Thus, the number of satellites can shed light on the process that limited the planet's growth.

The largest satellite that can be created in a gas-starved disk around a jovian-mass planet is roughly equal in size to Ganymede, Callisto, and Titan, the largest satellites of Jupiter and Saturn, $M\sim 10^{23}$ kg and $R\sim$ 2500 km.  Detailed simulation of the behavior of the circumplanetary disk and accretion within the disk is needed to determine if the $M_T\sim 2.5\times10^{-4} M_{pl}$ ratio would hold around much more massive planets.  Interestingly, {Ganymede, Callisto, and Titan} are large enough that the energy per unit mass deposited in an impact at escape velocity, $L_{imp} \sim v_{esc}^2/2 \sim 3.6 \times 10^6$ J/kg, is comparable to the latent heat of vaporization of ice.  Incredibly, this does not necessarily mean the satellites will melt upon formation; if the satellites form slowly, after the decay of short-lived radioisotopes, and on a long enough time scale to permit radiative cooling between accretional impacts, satellites of the sizes of Callisto and Titan can avoid global melting during formation \citep{cr, Barr:2010ab}.  (Of further interest, $L_{imp}$ for the Earth is comparable to the latent heat of vaporization for rock \cite{PS00}.)  It is possible that this energy balance may impose a hard upper limit to the size of icy satellites, irrespective of the mass of their parent planet.

It is important to note that although the gas-starved disk model readily explains the total masses of satellite systems and the maximum satellite mass, it is certainly possible that other processes may happen after the satellites are fully formed, which modify the architecture of the system.  It has been recently suggested that Saturn's satellites may have re-formed in the last 100 Myr, from a disk of material created from the collisions of a pre-existing satellite system \cite{Cuk2016}.  If this is the case, the total mass of satellite system may be primordial, but the masses of the individual satellites, and in particular, the mass of largest satellite, may reflect the local surface mass density profile of the disk and not a balance between growth and orbital decay.  

\subsection{Rings}
All four of the outer planets in our Solar System have rings.  The ring systems of Jupiter, Uranus, and Neptune have masses of about $10^{19}$ grams, roughly the mass of an icy moon 15 kilometers in radius.  These systems are thought to form via the disruption of very small inner satellites \citep{LukeRings}.

The bright, massive ring system of Saturn stands out as an oddity in both mass and composition.  It is difficult to estimate the mass of a ring system without an orbiting spacecraft, so reliable estimates of the mass of the rings have come about only recently, based on \emph{Cassini} observations.  Much of the mass is concentrated in the A and B rings, which together contain about $10^{22}$ to $10^{23}$ grams of icy material \citep{Robbins2010, Cuzzi2010}.  If all of the ring material were gathered into a single object, it would correspond to an icy satellite 150 to 300 km in radius, roughly the size of Saturn's moon Mimas.  This is a factor of 1000 to 10,000 larger than the mass of Jupiter's rings, and largely ruled out the idea that the rings had formed from the disruption of a small saturnian moon or a large comet \citep{LukeRings, CharnozLHB}.  

Estimates of the ring mass at Saturn suggested a new hypothesis for ring origin, which is intimately connected to the formation of Saturn's satellites in a gas-starved disk \citep{Canup2010}.  In the gas-starved disk, several generations of satellites are formed and lost to inward migration during the period of active inflow of gas to Saturn.  Canup (2010) \citep{Canup2010} proposes that Saturn's rings could represent the debris from a disrupted satellite that was ripped apart by tides.  If the satellite had fully differentiated into a rock core and ice-rich mantle, the icy mantle of the protosatellite would be stripped from the rock core and remain in orbit, with the rock core being absorbed by Saturn \citep{Canup2010}.  This could explain the ice-rich composition of the rings, and their mass.  The formation of a large ring system at Saturn could also potentially explain why Saturn has a number of small and mid-sized satellites, whereas Jupiter only has four relatively large moons \citep{Charnoz2011}.  If a hugely massive system of rings were formed early in Saturn's history, the small and mid-sized satellites (e.g., Mimas, Enceladus, Tethys, Dione, etc.) could have formed from ring debris that spread beyond Saturn's Roche limit \citep{Charnoz2011}.

The relationship between rings and moons is of particular interest at the present, in the wake of the discovery of a massive ring system around a companion planet to the roughly solar-mass star JWASP J140747.93-394542.6 \citep{Mamajek2015, RiederKenworthy2016}.  Oddities in the light curve of JWASP 1407 suggested the presence of a massive companion harboring a system of rings which contained gaps\citep{Mamajek2015}.  One way of opening gaps in a system of rings is with torques due to the presence of an embedded satellite; this led to the suggestion that JWASP 1407b may be orbited by a moon with mass $\sim 0.8 M_E$ \citep{Mamajek2015}.  Subsequent $N$-body simulations of the stability of possible ring systems suggest that J1407b, has a mass between 60 to 100 Jupiter masses.  If the ring system of J1407b indeed has an embedded satellite, $M_S/M_{pl} \sim 2$ to 4$\times 10^{-5}$, consistent with the values achievable in a gas-starved disk \citep{CW06}.  If the gas-starved disk model could be extrapolated to planetary masses consistent with J1407b, disruption of a primordial satellite could form the ring system.  However, the $N$-body simulations show that \emph{retrograde} ring systems are more stable at J1407b than prograde systems, which might rule out disruption of a regular satellite as a means of forming the rings.

\section{Other Means \label{sec:other-means}}
There are two additional theories of satellite formation, fission and capture, which were proposed as a means of forming our Solar System's satellites but ruled out by observational constraints.  It is worth discussing each of them briefly, because they could operate in exoplanetary systems.  These alternate theories were proposed for the formation of Earth's Moon, and a very detailed discussion of these may be found in Wood (1986) \cite{Wood1986}.  Here, I describe these theories qualitatively, because they may prove viable for exomoons discovered in the future, whose observed properties might be quite dissimilar from the satellites of our Solar System.

\subsection{Fission}
Fission has been considered for many years as a possible means of forming Earth's Moon.  The theory has many advantages in terms of explaining the similarities between the chemical composition and stable isotope ratios of the Moon and mantle of the Earth \cite{Wood1986}.  The theory has fallen out of favor in recent decades as models of terrestrial planet formation show that embryo collisions are common (e.g., \cite{Agnor1999}) and numerical impact simulations have explicitly demonstrated the injection of a lunar mass of material into orbit via embryo collisions.  

The fission hypothesis was most famously proposed by G. H. Darwin, son of Charles Darwin.  Darwin showed that if one put all the angular momentum of the Earth/Moon system back into the rotation of the Earth, the two-hour rotational period is comparable to the period of tidal forcing of the Earth by the Sun.  The commensurability between these two periods would cause the tides on the Earth to grow in amplitude, eventually causing some of the material to break off, forming the Moon \cite{Darwin1879}.  Fission by this mechanism was ruled out by Jeffreys (1930) \cite{Jeffreys1930}, who pointed out that tidal friction would prevent the tidal bulge from growing in amplitude.  Other variants of this theory invoke the formation of Earth's core (and concomitant change in its spin rate and moment of inertia) as a means of triggering fission \cite{Ringwood1960, Wise1963}.  

Another formulation of the fission theory is that the Earth is initially rotating so fast that it is dynamically unstable \cite{Darwin1880}.  This theory was ruled out as a viable means of forming Earth's Moon because the angular momentum required for fission due to pure rotational instability is a factor of 3 higher than the present value for the Earth/Moon system \cite{Stevenson1987}.  Hydrodynamical simulations of this process using simple polytropic equations of state for the Earth's core and mantle \cite{ DurisenScott1984} show that the material removed from the Earth forms an orbiting disk rather than an intact Moon.  This is different from Darwin's predictions, but somewhat more consistent with the modern view of the accretion of the Moon from a disk of material orbiting the Earth. 

\subsection{Capture}
Theories of the formation of planetary satellites by capture fall into two main categories: intact capture, in which the entire satellite ends up in orbit around the parent planet; and disintegrative capture, in which an existing body is disrupted and re-accretes in orbit around the parent body.  A main difficulty with any capture theory is to find a way to dissipate the excess orbital energy of bodies making a close approach to the planet (see \cite{KaulaHarris1975} for discussion).  A body entering the vicinity of a planet with a large relative velocity must dissipate enough energy to remain gravitationally bound to the planet after a single encounter.  

One method of removing excess energy is via gas drag \cite{Nakazawa1983, Pollack1979GasDrag}.  The protomoon passes through the atmosphere of a planet and sheds excess energy via fraction.  This requires the protomoon to enter the system on a trajectory where it will encounter the atmosphere at just the right angle to encounter enough gas, but also avoid merging with the planet, which is a highly unlikely event.  This mechanism seems to be more viable for gas giant planets with thick gaseous envelopes.  But it could work around a rocky planet with a thick atmosphere as well; gas drag was suggested as a possible means for forming Earth's Moon \cite{Nakazawa1983}, but was ruled out because there was no obvious means of removing the thick atmosphere of the protoearth.  Another suggested mechanism is disruptive capture, in which an object passes within the Roche limit of a planet and is broken into small pieces. If the encounter velocity is slow, material on the planet-facing side of the disrupted object can remain in orbit around the planet \cite{Opik1972, WoodMitler}.  This mechanism has been studied in only an approximate sense \cite{Wood1986} because early numerical simulations of this process showed that purely fluid objects (with no inherent strength) would simply not have enough time to break up during passage through the Earth's Roche limit \cite{MizunoBoss1985}.

Another form of intact capture, dynamical capture, has been shown to be a viable means of placing satellites in to orbit around a planet with a pre-existing satellite system, however \cite{Tsui1999, AgnorHamiltonTriton}.  Typically, dynamical capture is invoked as a means of capturing the numerous small satellites of the gas giants.  However, dynamical capture is thought to have formed Neptune's retrograde satellite Triton, thought to be a captured Kuiper Belt object.  The theory holds that a binary Kuiper Belt object interacted with Neptune's existing satellite system, leaving one member of the binary pair behind \cite{AgnorHamiltonTriton}.  The other member of the pair was ejected from the system, carrying with it the excess orbital energy. For this mechanism to work, the mass of the satellite system in place at Neptune is comparable to the mass of the satellite left behind \cite{AgnorHamiltonTriton}, so one may view this more as an ``exchange'' of satellite material rather than a true ``capture.'' 

\section{Dynamical Stability}
After a moon has formed, it will quickly evolve into a synchronous state \cite{MurrayDermott, BarnesOBrien2002, Meyer2010}, in which the satellite migrates either inward or outward so that its orbital mean motion $n=\Omega$, where $\Omega$ is the spin frequency of the planet.  Afterwards, the satellite continues to migrate outward to larger semi-major axes as the planet de-spins; if the planet-moon system is extremely close to the parent star, the moon can be stripped away by solar tides \cite{BarnesOBrien2002}.  The moon is in danger of being stripped if the semi-major axis of its planetary orbit exceeds the radius of the planet's Hill sphere, the region in which the planet's gravity dominates over that of the star:
\begin{equation}
R_H=a_p \bigg(\frac{M_p}{3M_*}\bigg)^{1/3},
\end{equation}
where $a_p$ is the semi-major axis of the planet's orbit around the star (which has mass $M_*$).  This provides a possible rationale for the lack of exomoon detections so far: as a planet migrates closer to its central star, the Hill sphere of the planet contracts.  Eventually, the moons are outside the planet's Hill sphere and can be lost to the star \cite{BarnesOBrien2002}. The timescale for this evolution depends on the rate of tidal energy dissipation in the planet \cite{BarnesOBrien2002},
\begin{equation}
T = \frac{2}{13} \frac{(\frac{4}{3} \pi \rho_p)^{5/3}}{\sqrt{G}} \bigg(\frac{(D_H a_*)^3}{3 M_*}\bigg)^{13/6} \bigg(\frac{Q_p}{3 k_{2,p}}\bigg)\bigg(\frac{M_{p}}{M_L}\bigg),
\end{equation}
where $M_*$ and $a_*$ are the mass and distance of the parent star, $\rho_p$ is the planet's density, $D_H=0.4895$ is the maximum planet-moon distance at which the moon remains bound to the planet when perturbed by solar gravitational forces \cite{Domingos2006}.  The values $Q_p$ and $k_{2,p}$ are the tidal quality factor and degree-2 Love number of the planet, which control the energy dissipation rate.  Because $T \varpropto M_L^{-1} a_*^{13/2}$, large moons orbiting planets close to their parent stars have short lifetimes; this may explain why there have been only a few possible detections of exomoons so far \citep{Bennett2014, Hippke}.

\section{Detection \label{sec:detection}}

Dozens of methods have been suggested for the detection of exomoons, and an excellent summary of many of these techniques may be found in Heller et al., (2014) \cite{HellerAbio2014}.  Here, I describe detection by three techniques that seem to be among the most promising in terms of characterization of exomoon/planet systems, and for constraining the prevalence of exomoons: (i) transit techniques, (ii) microlensing, and (iii) spectral/photometric techniques.  Figure \ref{fig:detection_meth} illustrates, qualitatively, how each of these three detection techniques is sensitive to different physical properties of the system, e.g., $M_S/M_{pl}$ and semi-major axis of the satellites.  Thus, each technique is sensitive to moons formed by different processes: microlensing can detect satellites even around small planets, spetroastrometric techniques favor large satellites with $M_S/M_{pl}$ with large semi-major axes, and transits favor large $M_S/M_{pl}$ with small semi-major axes.

\subsection{Transits}
The most well-developed exomoon detection techniques search for the gravitational and photometric effects of moons in transit data.  Photometric effects, such as a dip in the transit light curve from a moon during transit of its host planet, yield information about the ratio of the radius of the satellite to the radius of the star, $R_s/R_*$.  Gravitational effects, such as changes in the timing of transits, yield $M_s/M_{pl}$.  Light curves from \emph{COROT} and \emph{Kepler} provide an enormous data set in which to search for the effects of exomoons \cite{WeidnerHorne2010, Kipping2012-1, KippingKepler90}.  

Figure \ref{fig:wobble} illustrates movement of a planet-moon system around a central star.  A planet hosting a large moon (or family of moons) orbits around the center of mass of the system, rather than following a Keplerian orbit around the central star.  Thus, the length of time in between transits of the planet across the star is different from what would be predicted for a solo planet on a Keplerian orbit \cite{Sartoretti1999}.  This effect is called ``transit timing variation,'' or TTV.  The magnitude of transit timing variations is proportional to the product $(a_s/a_{pl})(M_s/M_{pl})$ \cite{Sartoretti1999}, where $a_s$ is the satellite's orbital semi-major axis around the system center of mass, and $a_{pl}$ is the planet's semi-major axis about the central star.  

The non-Keplerian nature of the host planet's orbit also changes the effective velocity of the planet/moon system across the central star during the transit \cite{Kipping2009a}.  This effect can change the duration of the transit, and is known as ``transit duration variations'' or TDV.  The magnitude of transit duration variations are proportional to $(a_s/a_{pl})(M_s/M_{pl})(P_{pl}/P_{s})$, where $P_{pl}$ is the period of the planet's orbit around the central star, and $P_{s}$ is the satellite's orbital period around the planet \cite{Kipping2009b}.  

Detection of both effects, although technically challenging, permits exomoon signals to be distinguished from other effects such as multiple planets \cite{KippingPOS}.  The ratio of the magnitude of TTV to that of TDV yields a separate estimate of the satellite's semi-major axis and mass \cite{KippingPOS}.  Because TTV scales as $M_s a_s$ and TDV scales as $M_s a_s^{-1/2}$, moons close to the planet have large TDV but smaller TTV.  The converse is true, as well: moons far from their host planet have small TDV but large TTV.  Thus, both effects can be detected only for satellites at a certain intermediate distance from their host planet \cite{KippingPOS}.  The most thorough application of TTV and TDV to the search for exomoons has been performed by the Hunt for Exomoons with \emph{Kepler} (HEK) team \cite{Kipping2012-1}.  Folding in uncertainties in planetary parameters, and signal-to-noise considerations, the HEK team estimate that they can detect exomoons with masses $\gtrsim 0.1$ Earth masses \cite{Kipping2012-1}.   

\subsection{Microlensing}
Gravitational microlensing is now a well-known technique for detecting extrasolar planets.  Gravitational lensing occurs when light from the source star is bent toward the observer by a large mass in between the star and observer.  The light from the source is distorted into a ring; the radius of the ring is sensitive to the mass of the lensing object, and the distance between the lens and the source.  In a microlensing event, the mass of the lensing object is not large enough to form a detectable ring -- instead, the brightness of the lensing object increases momentarily.  Planets and moons orbiting the source star can be detected by examining how the brightness changes as a function of time: if there are features in the signal that cannot be explained by the presence of a planet alone, this provides evidence for the presence of a moon \citep{Liebig}.  Giant stars are most efficient lenses, but have a tendency to smooth out the small variations in brightness that might indicate the presence of a moon.  The most effective means of searching for exomoons is to monitor dwarf stars equal in size or smaller than the Sun from a space-based observatory \citep{Liebig}.  Microlensing techniques have about a 1-in-3 chance of detecting the Earth-Moon system in orbit around a sun-like star at a distance of 8 kiloparsecs. 

Microlensing has the potential to detect smaller $M_S/M_{pl}$ than the transit technique, with presently available instrumentation.  It may be some time before transit or other techniques are able to find planet/moon systems that resemble those in our Solar System.  The disadvantage of microlensing is that repeat observations are not possible, and so possibilities of more detailed characterization of the planet-moon system are limited.  However, numerous detections would provide statistics on the likelihood of moon systems around extrasolar planets in general.

\subsection{Spectral/Photometric Techniques}
A newly proposed technique, ``spectroastrometric'' detection, offers the possibility of both detection and detailed characterization of exomoons \citep{Agol2015}.  The technique takes advantage of the differences in the light emissions from a planet and moon.  The planet/moon system is directly imaged through the use of a coronagraph which suppress the light from the star.  The planet and moon are assumed to be emitting energy with different spectra: thus, the location of the centroid of the light detected from the planet/star system will change position depending on wavelength \citep{Agol2015}.  The change in centroid location provides information about the physical separation between the planet and moon.  Given sufficient observation time, the spectra of the planet and moon could potentially be disentangled \citep{Agol2015}, allowing for compositional characterization of the system, yielding much more information than just the mass and radius of both bodies. 

The technique requires a large space-based telescope with a coronagraph and sufficient spectral coverage and resolution to detect absorption bands from chemical species such as water and methane that are likely to be found on parent planet \citep{Agol2015}.  Like microlensing and transits, the viability of the technique has only been characterized for very specific satellite systems: the Earth/Moon system, and the potentially unrealistic scenario of the Earth orbiting a Jupiter-mass planet \citep{Agol2015}.  Similar to the transit technique, detection and characterization is favored for large moon-to-planet mass ratios.  But unlike transits, this technique can function well for exomoons orbiting far from their parent planet.  

\section{Discussion: How to Make a Large Exomoon \label{sec:extrapolation}}
\subsection{In a Collision}
It is presently difficult to speculate about the maximum $M_s/M_{pl}$ that could be created in a collision between two solid planets (i.e., planets without a gaseous envelope).  This is simply because so many of the numerical simulations of planetary collisions have been performed to explain the formation of Earth's Moon (e.g., \cite{CanupAsphaug, Canup2004}), or at conditions not conducive to the formation of moons at all (e.g., \cite{MarcusRocky2009, MarcusIcy2010}).  Thus, the maximum value of $M_o$, the orbiting disk mass, that might be achieved in a collision between planets of a given size, composition, and size ratio, is not well constrained.  

Even if one knew the maximum $M_o$, the extent to which the impact-generated disk of material falls back onto the planet, versus spreads beyond the Roche limit to accrete into a satellite (or satellites), has only been studied for rock-rich disks appropriate for the Moon-forming impact.  Very little is known about how the Roche-interior disk might behave if it had a higher metal content, or contained significant amounts of water.  These situations could easily arise in other Solar Systems where the terrestrial planets might have compositions very different from the Earth.

In spite of these uncertainties, some progress has been made on the number of collisional satellites that might be expected in other Solar Systems.  Elser et al., (2011) \cite{Elser2011} used the scaling relationships between disk mass and collision conditions based on Moon-forming impacts, coupled with impact histories from $N$-body simulations of terrestrial planet formation \cite{Morishima2010} to speculate about the sizes and prevalence of moons around terrestrial planets, in general.  Their work suggests that roughly 2 to 25\% of terrestrial planets should harbor a large moon.  

Additional hydrocode simulations of impacts are needed to provide estimates of the disk masses created in impacts unlike the Moon-forming impact, and for planets of differing iron and water ice content.  One need only look at the differences between the Moon-forming and Pluto-Charon-forming impacts to see the range of different outcomes that can occur in a planetary collision.  Although the mass of the impact-generated disk can serve as a hard upper limit to the mass of moon created in an impact \cite{Elser2011}, simulations of the evolution of the disk and accretion of impact-generated debris are needed to determine whether a large disk would accrete into a single satellite, or several smaller bodies.  

\subsection{Around a Giant Planet}
Similar to planetary collisions, accretion of satellites around gas giant planets has been modeled in exquisite detail, and with great success, for conditions relevant to our Solar System.  

Although coupled $N$-body and disk simulations of the type performed by Canup \& Ward (2006) \cite{CW06} have only been done for Jupiter-mass planets, disk models have been constructed for super-jovian planets at varying distances from a sun-like star \cite{Heller2015a, Heller2015b}.  These studies explore disk conditions at super-jovian planets, at a variety of distances away from their parent star, and find that super-jovian planets that form more than 5 AU from a sun-like star can harbor moons up to Mars' mass (about 4$\times$ the mass of Ganymede) \cite{Heller2015a, Heller2015b}.  Varying the stellar distance has the effect of changing the relative heating rates from insolation and gravitational contraction of the planet, as well as the composition of material pulled into the circumplanetary disk \citep{Heller2015a, Heller2015b}.  

A more massive planet will liberate more gravitational potential energy as it contracts, yielding a hotter disk.  Moreover, solid material will be pulled into planetary orbit, potentially yielding more massive satellites.  An additional consideration is that stars with higher or lower metallicity might have a different mass ratio of solids to gas, which would affect the amount of solids available to create satellites.  Another question is how the gas-starved disk model might be affected by the presence of deuterium burning.  The additional planetary luminosity from deuterium burning might make the disks too hot to accrete ice, or perhaps even rock, during the early stages of the planet's growth.  But if the planet is able to cool sufficiently to allow solids to remain stable in the disk, satellites may still form.

Finally, it is worth noting that the outer planets in our Solar System, and also super-jovian planets large enough to burn deuterium, could form via the direct collapse of a portion of the protoplanetary disk, rather than core accretion \citep{Boss97}.  Unfortunately, there are no detailed models of satellite accretion in the context of this formation mechanism.  This could present a fruitful avenue for future study.

\subsection{Processes that Did Not Operate in our Solar System}
It is certainly possible that large exomoons could form via fission or capture, which were ruled out as means of forming moons in our Solar System.  Pure fission (as opposed to impact-induced fission) of a terrestrial planet or large icy body has not been studied using modern numerical techniques, so it is difficult to speculate about how it might operate, or the conditions that might lead to its occurrence.  The same holds for disruptive capture, which was ruled out as a means of forming any of the Solar System's satellites, but a basic physical plausibility argument has not been constructed using modern numerical techniques.

Some preliminary work has been done to determine the likelihood of capture of a large exomoon at a gas giant via binary exchange, the type of capture mechanism thought to have resulted in Triton's capture at Neptune \cite{Williams2013}.  This mechanism relies upon the existence of large binary objects with rapid rotational velocities (which would theoretically have formed in a collision).  These binary systems encounter a giant planet, leaving one member of the binary system intact, and on a highly elliptical orbit.  The orbits may then become circularized through the Kozai mechanism \citep{PorterGrundy2011}, leaving the system in a dynamically stable state.

\section{Prospects for Future Work}

The past few decades have seen a renaissance in our understanding of the formation and evolution of planetary satellites.  We have learned that the masses, compositions, and orbits of the moons at each Solar System object can yield clues about how that planet formed: the Earth and other terrestrial planets suffered frequent violent collisions; the giant planets formed circumplanetary disks and multiple generations of satellites during the lifetime of the solar nebula \cite{CW06}.  

When we consider extrapolating our Solar System knowledge to speculate about moon formation in other Solar Systems, we quickly bump up against the limitations of Solar System studies.  Many of the modern theories of satellite formation are tightly tailored to create Solar System planet/moon families, and precious little effort has been spent considering how these processes might change in other systems.  

The following are a series of questions that could be addressed in future studies, which would help us understand how detectable exomoons could form, and how common large exomoons might be:
\begin{itemize}
\item Does the $M_{S}/M_{pl}$ predicted by the Canup \& Ward (2006) \cite{CW06} model hold for gas giants with mass, composition, stellar metallicity, stellar type, and planet-star distance vastly different from the conditions of jovian accretion?
\item How does deuterium burning in the host planet affect the formation of satellites at gas giant planets?
\item How does the formation of planetary ring systems scale with planetary mass?
\item What is the maximum mass of orbiting disk, and single satellite, that can be created in an impact between rock/metal, ice/rock, and ice/rock/metal planets?
\item Do the proportions of water/metal/rock in an impact-generated disk change the efficiency with which material is accreted into satellites versus lost to the host planet?
\item Are there fission or capture scenarios that can create satellites large enough to be observed with \emph{Kepler}?
\end{itemize}

\section{Acknowledgements}  I thank the organizers of the Space Telescope Science Institute Habitable Worlds through Space and Time conference of 2014, in particular, John Debes, for inviting me to give a talk which motivated the writing of this paper.  I also thank Ramon Brasser for discussions regarding accretion processes in impact-generated disks and the gas-starved disk.

\clearpage
\begin{table}[ht!]
\tbl{Properties of the Major Satellites in our Solar System and Their Modes of Origin}
{\begin{tabular}[l]{@{}llllllll}\toprule
 Planet & Satellite & Orbital  & Orbital & Density  & $H_{sats}/H_{tot}$$^d$ & $M_s/M_{pl}$ & Origin\\
  	&	& Radius$^a$ & Character$^b$  & Ratio$^c$ &  & & Mode\\
\colrule	
Earth & Moon & 60 & RT & 0.61 & 0.83 & 0.012 & Impact \\
\hline
Mars &  &  & & & $10^{-6}$ & $1.8 \times 10^{-8}$ & ? \\
  & Phobos & 2.8 & RT & 0.48 &   & $1.6 \times 10^{-8}$ &    \\
  & Deimos & 6.9 & RT & 0.44 & & $2.3 \times 10^{-9}$ &  \\
\hline
Pluto &  &  & & & 0.99 & 0.12& Impact \\
	& Charon & 16 & RT & 0.92 &  &  0.12 &\\
	& Nix$^e$ & 36 & RT &  & & $3.4\times10^{-6}$ & \\
	& Styx$^e$ & 41 & RT &  & & $5.8\times10^{-7}$ &  \\
	& Kerberos$^e$ & 49 & RT &  & &  $1.3 \times 10^{-6}$ & \\
	& Hydra$^e$ & 54 & RT &  & &  $3.7 \times 10^{-6}$ & \\
\hline
Jupiter&  &  & & & 0.011 &  $2.06\times10^{-4}$& Co-Accretion \\
	& Io & 6.0 & RT & 2.52 & & $4.7\times10^{-5}$ &\\
	& Europa & 9.6 & RT & 2.27 & & $2.5\times10^{-5}$ &\\
	& Ganymede & 15.3 & RT & 1.46 & & $7.8\times10^{-5}$ & \\
	& Callisto & 26.9 & RT & 1.38 & & $5.6\times10^{-5}$ &\\
\hline	
Saturn & &  & & & 0.014 & $2.39 \times 10^{-4}$&Co-Accretion \\
	& Mimas & 3.2 & RT & 1.67 &  & $6.7 \times 10^{-8}$ & \\
	& Enceladus & 4.1 & RT & 2.34 &  & $1.9 \times 10^{-7}$ & \\
	&Tethys & 5.1 & RT & 1.43 &  & $  1.1\times 10^{-6}$ & \\
	&Dione & 6.5 & RT & 2.15 &  & $1.9  \times 10^{-6}$ &  \\
	& Rhea & 9.0 & RT & 1.81 &  & $ 4.4\times 10^{-6}$ &  \\
	& Titan & 21 & RT & 2.73 &  & $ 2.3 \times 10^{-4}$ &  \\
	&Hyperion & 2.5 & RT & 0.80 &  & $ 9.9\times 10^{-9}$ &  \\
	& Iapetus & 6.1 & RT & 1.59 &  & $3.2\times 10^{-6}$ & \\
\hline
Uranus&   &  & & & 0.011 &  $1.05\times10^{-4}$ & Co-Accretion \\
	 & Miranda & 5.1 & RT & 0.94 &  & $7.6\times 10^{-7}$ &\\
	 & Ariel & 7.5 & RT & 1.31 &  & $1.6\times 10^{-5}$ & \\
	 & Umbriel & 10 & RT & 1.10 &  & $1.3\times 10^{-5}$ & \\
	 & Titania & 17 & RT & 1.34 &  & $4.1\times 10^{-5}$ & \\
	 & Oberon & 23 & RT & 1.28 &  & $3.5\times 10^{-5}$ &\\
\hline
Neptune	&  &  & & & 0.02 & $2.1 \times 10^{-4}$ & Capture \\
	& Triton & 14 & $i \sim 157^{\circ}$$^e$ & 1.25  & & $2.1\times 10^{-4}$ & \\
	& Nereid & 224 & $i \sim 7^{\circ}$ & &  &   $2.6\times 10^{-7}$ & \\
\botrule
\end{tabular}}
\tabnote{$^{\rm a}$Approximate orbital distance scaled by planetary radius.}
\tabnote{$^{\rm b}$RT=regular, tidally evolved \cite{StevensonSatsBook}.}
\tabnote{$^{\rm c}$Satellite density divided by the mean density of the primary planet.}
\tabnote{$^{\rm d}$Ratio of angular momentum in satellites to total system angular momentum, after \cite{MacDonald1966}.}
\tabnote{$^{\rm e}$Upper limits from \cite{Brozovic}.}
\label{table:satellites}
\end{table}

\clearpage

\clearpage

\clearpage

\clearpage

\clearpage
\begin{figure}
\includegraphics[width=80mm]{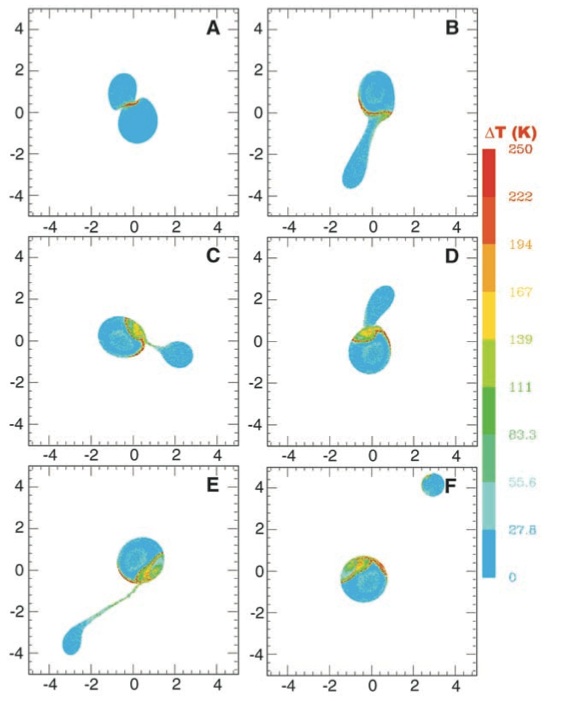}
\caption{Reproduced with permission from Canup (2005) \cite{Canup2005}.   Time-evolution of a smoothed particle hydrodynamics simulation of a potential Pluto-Charon-forming impact between two pure serpentine objects (run20 in Table 1 of Canup (2005)).  The results are shown at times (A) $t=0.9$, (B) 3.2, (C) 5.9, (D) 7.5, (E) 11.2, and (F) 27.5 hours.  Colors indicate the change in temperature from the initial condition, with red indicating $\Delta T=$ 250 K, and blue indicating no temperature change.  Distances are shown in units of $10^3$ km.  The objects collide with an angle of 73$^{\circ}$, at $v_{imp}=v_{esc}$.  The final state is a single intact moon with $M_s/M_{pl}=0.12$.  \label{fig:Canup2005}}
\end{figure}

\clearpage
\begin{figure}
\includegraphics[width=140mm]{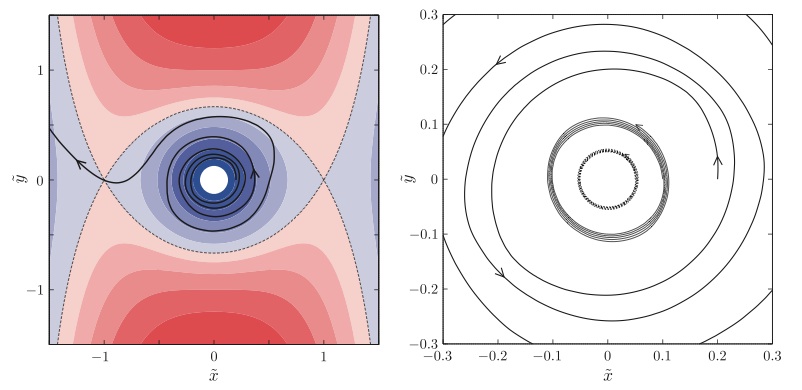}
\caption{Reproduced, with permission from Figure 2 of Tanigawa et al., (2012) \cite{TanigawaPaper1}. (left) Colors indicate values of the potential relative to Lagrange points L$_1$ and L$_2$.  The streamline starting from $r=0.2R_H$ ($r \approx 150 R_J$ for a Jupiter-mass planet) indicates particle trajectories in the midplane of a circumplanetary disk.  Dimensions are scaled by planetary Hill radii.  (right) Same as left panel, showing particle streamlines originating at $r=0.2R_H$, $0.1R_H$ ($75 R_J$), and $0.05R_H$ ($37 R_J$). \label{fig:Tanigawa}}
\end{figure}

\clearpage
\begin{figure}
\includegraphics[width=80mm]{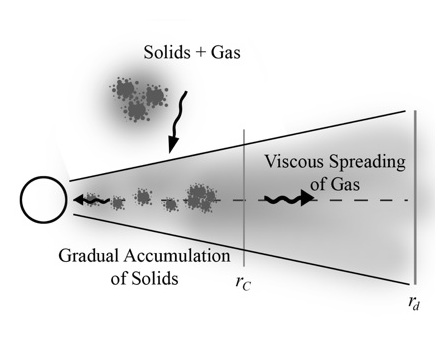}
\caption{Schematic of the gas-starved disk, based on models of Canup \& Ward (2002, 2006) \cite{CW02, CW06}. Ice/rock particles + gas from the solar nebula deliver mass to $r < r_C$. Gas spreads beyond $r_C$ and on to the planet. Solids in the midplane grow large enough to de-couple from gas, build up in the midplane and accrete into satellites. Satellite growth times and disk temperatures are controlled by the inflow rate.\label{fig:gsd}}
\end{figure}

\clearpage
\begin{figure}
\includegraphics[width=100mm]{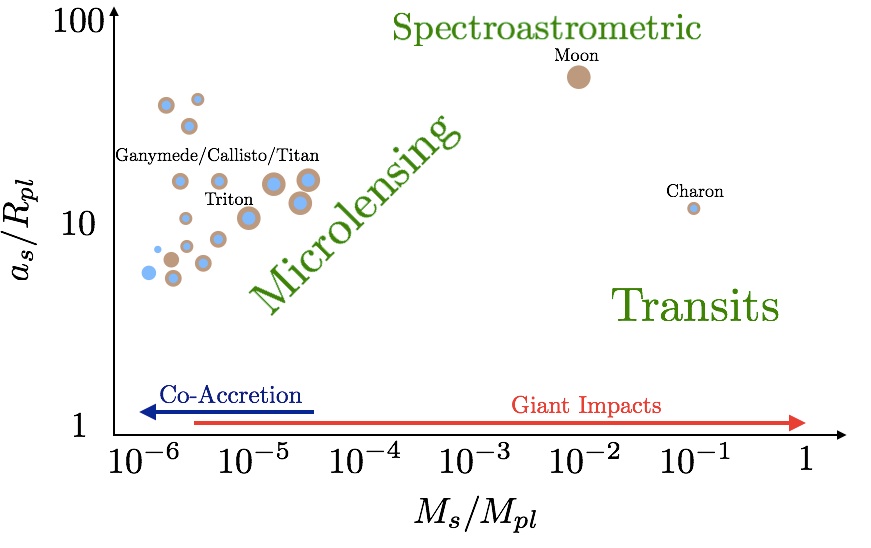}
\caption{Schematic depicting the power of microlensing, spectroastrometric, and transit exomoon detection and characterization methods, as a function of satellite-to-planet mass ratio, and satellite semi-major axis.  Approximate location of the Solar System's satellites are shown as dots: brown/blue dots indicate the mixed ice/rock satellites of the outer planets and Charon, the Moon is represented by a brown dot.  Giant impacts have the potential to form the large $M_S/M_{pl}$ required to observe exomoons via transits and spectroastrometric methods.  Microlensing has the potential to observe smaller satellites.  What is \emph{not} depicted on this graph is the difficulty of detecting the \emph{host planet} to begin with.  All of our Solar System's planets are barely detectable using present instrumentation: there is essentially no chance of detecting a dwarf planet like Pluto using any present or planned telescopes.  \label{fig:detection_meth}}
\end{figure}

\clearpage
\begin{figure}
\includegraphics[width=80mm]{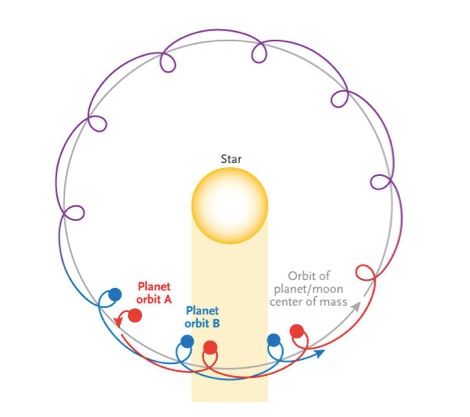}
\caption{Reproduced, with permission, from Kipping 2014 \cite{KippingPOS}.  Schematic illustration of the effect of an orbiting moon on the timing and duration of planetary transits.  The center of mass of the planet/satellite system follows a Keplerian orbit around the central star (gray line), but the location of planet itself (red/blue) deviates from a Keplerian orbit (red/blue).  This causes changes in the timing and duration of transits relative to those predicted for a planet alone. \label{fig:wobble}}
%
\end{figure}


\begin{thebibliography}{100}
\newcommand{\noopsort}[1]{}
\newcommand{\printfirst}[2]{#1}
\newcommand{\singleletter}[1]{#1}
\newcommand{\switchargs}[2]{#2#1}
\providecommand{\url}[1]{\normalfont{#1}}
\providecommand{\urlprefix}{Available at }

\bibitem{Agnor1999}
C.B. {Agnor}, R.M. {Canup}, and H.F. {Levison}, Icarus 142 (1999), pp.
  219--237.

\bibitem{Chambers2004}
J.E. Chambers, Earth Planet. Sci. Lett. 223 (2004), pp. 241--252.

\bibitem{cr}
A.C. {Barr} and R.M. {Canup}, Icarus 198 (2008), pp. 163--177.

\bibitem{Barr:2010ab}
A.C. Barr, R.I. Citron, and R.M. Canup, Icarus 209 (2010), pp. 858--862.

\bibitem{Sasaki2010}
T.~Sasaki, G.R. Stewart, and S.~Ida, The Astrophysical Journal 714 (2010), p.
  1052.

\bibitem{Reynolds1987}
R.T. {Reynolds}, C.P. {McKay}, and J.F. {Kasting}, Advances in Space Research 7
  (1987), pp. 125--132.

\bibitem{Williams1997}
D.M. {Williams}, J.F. {Kasting}, and R.A. {Wade}, Nature 385 (1997), pp.
  234--236.

\bibitem{Kaltenegger2010}
L.~{Kaltenegger}, Astrophys. J. 712 (2010), pp. L125--L130.

\bibitem{Heller2012}
R.~Heller, Astron. Astrophys. 545 (2012), p.~L8.

\bibitem{HellerBarnes2013}
R.~{Heller} and R.~{Barnes}, Astrobiology 13 (2013), pp. 18--46.

\bibitem{Tarter2007}
J.C. Tarter, P.R. Backus, R.L. Mancinelli, J.M. Aurnou, D.E. Backman, G.S.
  Basri, A.P. Boss, A.~Clarke, D.~Deming, L.R. Doyle, \emph{et~al.},
  Astrobiology 7 (2007), pp. 30--65.

\bibitem{Burns1978}
J.A. Burns, Vistas in Astronomy 22 (1978), pp. 193--210.

\bibitem{HartmannDavis}
W.K. {Hartmann} and D.R. {Davis}, Icarus 24 (1975), pp. 504--515.

\bibitem{CameronWard}
A.G.W. {Cameron} and W.R. {Ward}, \emph{{The origin of the Moon}}, in
  \emph{Lunar and Planetary Science Conference Proceedings}, Vol.~7, Apr.,
  1976, pp. 120--122.

\bibitem{CanupAsphaug}
R.M. {Canup} and E.~{Asphaug}, Nature 412 (2001), pp. 708--712.

\bibitem{Canup2004}
R.M. {Canup}, Icarus 168 (2004), pp. 433--456.

\bibitem{Canup2005}
R.M. {Canup}, Science 307 (2005), pp. 546--550.

\bibitem{Canup2011}
R.M. Canup, The Astronomical Journal 141 (2011), p.~35.

\bibitem{StewartCuk2012}
M.~{{\'C}uk} and S.T. {Stewart}, Science 338 (2012), pp. 1047--1052.

\bibitem{Canup2012}
R.M. {Canup}, Science 338 (2012), pp. 1052--1055.

\bibitem{Chambers2013}
J.E. {Chambers}, Icarus 224 (2013), pp. 43--56.

\bibitem{OgiharaIda2009}
M.~Ogihara and S.~Ida, The Astrophysical Journal 699 (2009), pp. 824--838.

\bibitem{Morishima2010}
R.~Morishima, J.~Stadel, and B.~Moore, Icarus 207 (2010), pp. 517--535.

\bibitem{Elser2011}
S.~Elser, B.~Moore, J.~Stadel, and R.~Morishima, Icarus 214 (2011), pp.
  357--365.

\bibitem{PollackConsolmagnoSaturnBook}
J.B. {Pollack} and G.~{Consolmagno}, \emph{{Origin and evolution of the Saturn
  system}}, in \emph{Saturn}, University of Arizona Press, Tucson, AZ, 1984,
  pp. 811--866.

\bibitem{Coradini1989}
A.~{Coradini}, P.~{Cerroni}, G.~{Magni}, and C.~{Federico}, \emph{{Formation of
  the satellites of the outer solar system: Sources of their atmospheres}}, in
  \emph{Origin and Evolution of Planetary and Satellite Atmospheres},
  University of Arizona Press, Tucson, AZ, 1989, pp. 723--762.

\bibitem{CW02}
R.M. {Canup} and W.R. {Ward}, Astron. J 124 (2002), pp. 3404--3423.

\bibitem{ME03}
I.~{Mosqueira} and P.R. {Estrada}, Icarus 163 (2003), pp. 198--231.

\bibitem{WardCanup2010}
W.R. Ward and R.M. Canup, The Astronomical Journal 140 (2010), p. 1168.

\bibitem{CW06}
R.M. {Canup} and W.R. {Ward}, Nature 441 (2006), pp. 834--839.

\bibitem{Heller2015a}
R.~Heller and R.~Pudritz, The Astrophysical Journal 806 (2015), p. 181.

\bibitem{Heller2015b}
R.~Heller and R.~Pudritz, Astronomy \& Astrophysics 578 (2015), p. A19.

\bibitem{Kipping2012-1}
D.M. Kipping, G.{\'A}. Bakos, L.~Buchhave, D.~Nesvorn{\`y}, and A.~Schmitt, The
  Astrophysical Journal 750 (2012), p. 115.

\bibitem{Heller2014Detection}
R.~{Heller}, Astrophys. J. 787 (2014), 14.

\bibitem{Hippke}
M.~{Hippke}, Astrophys. J. 806 (2015), 51.

\bibitem{LewisPulsar}
K.M. Lewis, P.D. Sackett, and R.A. Mardling, ApJL 685 (2008), pp. L153--L156.

\bibitem{Noyola}
J.P. Noyola, S.~Satyal, and Z.E. Musielak, Astrophys. J. 791 (2014), p.~25.

\bibitem{Liebig}
C.~Liebig and J.~Wambsganss, aa 520 (2010), p. A68.

\bibitem{SenguptaMarley}
S.~Sengupta and M.S. Marley, The Astrophysical Journal 824 (2016), p.~76.

\bibitem{Mamajek2015}
M.A. Kenworthy and E.E. Mamajek, The Astrophysical Journal 800 (2015), p. 126.

\bibitem{Kipping2009b}
D.M. {Kipping}, Monthly Notices of the Royal Astronomical Society 396 (2009),
  pp. 1797--1804.

\bibitem{Sartoretti1999}
P.~Sartoretti and J.~Schneider, Astronomy and Astrophysics Supplement Series
  134 (1999), pp. 553--560.

\bibitem{Szabo2006}
G.M. {Szab{\'o}}, K.~{Szatm{\'a}ry}, Z.~{Div{\'e}ki}, and A.~{Simon}, Astron.
  Astrophys. 450 (2006), pp. 395--398.

\bibitem{Kipping2009a}
D.M. {Kipping}, Monthly Notices of the Royal Astronomical Society 392 (2009),
  pp. 181--189.

\bibitem{Domingos2006}
R.C. Domingos, O.C. Winter, and T.~Tokoyama, Monthly Notices of the Royal
  Astronomical Society 373 (2006), pp. 1227--1234.

\bibitem{Namouni2010}
F.~{Namouni}, Astrophysical Journal 719 (2010), pp. L145--L147.

\bibitem{BarnesOBrien2002}
J.W. Barnes and D.P. O'Brien, Astrophys. J. 575 (2002), pp. 1087--1093.

\bibitem{Nesvorny2007}
D.~Nesvorn{\`y}, D.~Vokrouhlick{\`y}, and A.~Morbidelli, The Astronomical
  Journal 133 (2007), p. 1962.

\bibitem{StevensonSatsBook}
D.J. {Stevenson}, A.W. {Harris}, and J.L. {Lunine}, \emph{{Origins of
  Satellites}}, in \emph{Satellites}, University of Arizona Press, Tucson, AZ,
  1986, pp. 39--88.

\bibitem{BodmanKIC2016}
E.H. Bodman and A.~Quillen, The Astrophysical Journal Letters 819 (2016), p.
  L34.

\bibitem{DobroPlutoBook}
A.R. {Dobrovolskis}, S.W. {Peale}, and A.W. {Harris}, \emph{{Dynamics of the
  {P}luto-{C}haron Binary}}, in \emph{Pluto and Charon}, University of Arizona
  Press, Tucson, AZ, 1997, pp. 159--190.

\bibitem{McKinnon1984}
W.B. McKinnon, Nature 311 (1984), pp. 355--358.

\bibitem{GoldreichNeptune}
P.~{Goldreich}, N.~{Murray}, P.Y. {Longaretti}, and D.~{Banfield}, Science 245
  (1989), pp. 500--504.

\bibitem{AgnorHamiltonTriton}
C.B. Agnor and D.P. Hamilton, Nature 441 (2006), pp. 192--194.

\bibitem{WolszczanFrail}
A.~Wolszczan and D.A. Frail, Nature 355 (1992), pp. 145--147.

\bibitem{Morby2012}
A.~Morbidelli, J.I. Lunine, D.P. O'Brien, S.N. Raymond, and K.J. Walsh, Ann.
  Rev. Earth Planet Sci. 40 (2012), pp. 251--275.

\bibitem{AsphaugReufer2014}
E.~Asphaug and A.~Reufer, Nature Geoscience 7 (2014), pp. 564--568.

\bibitem{AlemiStevenson}
A.~{Alemi} and D.~{Stevenson}, \emph{{Why Venus has No Moon}}, in
  \emph{AAS/Division for Planetary Sciences Meeting Abstracts \#38}, Bulletin
  of the American Astronomical Society, Vol.~38, Sep., 2006, p. 491.

\bibitem{Marinova2008}
M.M. Marinova, O.~Aharonson, and E.~Asphaug, Nature 453 (2008), pp. 1216--1219.

\bibitem{BurnsMars}
J.A. {Burns}, \emph{{Contradictory clues as to the origin of the Martian
  moons}}, in \emph{Mars}, University of Arizona Press, Tucson, AZ, 1992, pp.
  1283--1301.

\bibitem{Peale2007}
S.J. {Peale}, \emph{{The origin of the natural satellites}}, in \emph{Treatise
  on Geophysics}, Vol.~10, Elsevier, Amsterdam, 2007, pp. 456--508.

\bibitem{Craddock2011}
R.A. Craddock, Icarus 211 (2011), pp. 1150--1161.

\bibitem{CitronMarsMoons}
R.I. Citron, H.~Genda, and S.~Ida, Icarus 252 (2015), pp. 334--338.

\bibitem{KenyonBromley2013}
S.J. Kenyon and B.C. Bromley, Astrophys. J. 147 (2014), p.~8.

\bibitem{BenzMoon}
W.~Benz, A.G.W. Cameron, and H.J. Melosh, Icarus 81 (1989), pp. 113--131.

\bibitem{ReuferMoon2012}
A.~Reufer, M.M.M. {Meier}, W.~Benz, and R.~Wieler, Icarus 221 (2012), pp.
  296--299.

\bibitem{Meier2014}
M.M.M. {Meier}, A.~Reufer, and R.~Wieler, Icarus 242 (2014), pp. 316--328.

\bibitem{Hyodo}
R.~Hyodo, K.~Ohtsuki, and T.~Takeda, Astrophys. J. 799 (2015), p.~40.

\bibitem{moon_review}
A.C. Barr, J. Geophys. Res. in press.

\bibitem{Monaghan1992}
J.J. {Monaghan}, Ann. Rev. Astron. Astrophys. 30 (1992), pp. 543--574.

\bibitem{BenzMoonPaper1}
W.~Benz, W.L. Slattery, and A.G.W. Cameron, Icarus 66 (1986), pp. 515--535.

\bibitem{WadaNorman2001}
K.~{Wada} and C.A. {Norman}, Astrophys. J. 547 (2001), pp. 172--186.

\bibitem{Wada2006}
K.~Wada, E.~Kokubo, and J.~Makino, The Astrophysical Journal 638 (2006), p.
  1180.

\bibitem{McGlaun}
J.M. {McGlaun}, S.L. {Thompson}, and M.G. {Elrick}, Int. J. Imp. Eng. 10
  (1990), pp. 351--360.

\bibitem{Crawford2006}
D.A. {Crawford}, P.A. {Taylor}, R.L. {Bell}, and E.S. {Hertel}, Russ. J. Phys.
  Chem. B 25 (2006).

\bibitem{m-moon}
R.M. {Canup}, A.C. {Barr}, and D.A. {Crawford}, Icarus 222 (2013), pp.
  200--219.

\bibitem{IdaCanupStewart}
S.~{Ida}, R.M. {Canup}, and G.R. {Stewart}, Nature 389 (1997).

\bibitem{Kokubo2000}
E.~{Kokubo}, S.~{Ida}, and J.~{Makino}, Icarus 148 (2000), pp. 419--436.

\bibitem{ThompsonStevenson}
C.~{Thompson} and D.J. {Stevenson}, Astrophys. J. 333 (1988), pp. 452--481.

\bibitem{Ward2011}
W.R. Ward, Astrophys. J. 744 (2012), p. 140.

\bibitem{SalmonCanup}
J.~{Salmon} and R.M. {Canup}, Astrophys. J. 760 (2012), p.~83.

\bibitem{SalmonCanup2014}
J.~{Salmon} and R.M. {Canup}, Phil. Trans. R. Soc. A 372 (2014), p. 20130256.

\bibitem{Canup2008}
R.M. Canup, Icarus 196 (2008), pp. 518--538.

\bibitem{Leinhardt2012I}
Z.M. {Leinhardt} and S.T. {Stewart}, Astrophys. J. 745 (2012), 79.

\bibitem{GH}
D.E. {Gault} and E.D. {Heitowit}, \emph{{The Partition of Energy for
  Hypervelocity Impact Craters formed in Rock (NASA Technical Report Number
  NASA-TM-X-57428)}}, in \emph{Proceedings of the Sixth Hypervelocity Impact
  Symposium}, 1963, pp. 419--456.

\bibitem{MarcusRocky2009}
R.A. Marcus, S.T. Stewart, D.~Sasselov, and L.~Hernquist, The Astrophysical
  Journal Letters 700 (2009), p. L118.

\bibitem{MarcusIcy2010}
R.A. Marcus, D.~Sasselov, S.T. Stewart, and L.~Hernquist, Astrophysical Journal
  Letters 719 (2010), pp. L45--L49.

\bibitem{Genda2012}
H.~{Genda}, E.~{Kokubo}, and S.~{Ida}, Astrophys. J. 744 (2012), pp. 137--145.

\bibitem{TakedaIda}
T.~{Takeda} and S.~{Ida}, Astrophys. J. 560 (2001), pp. 514--533.

\bibitem{CridaCharnoz}
A.~Crida and S.~Charnoz, Science 338 (2012), pp. 1196--1199.

\bibitem{WardCameron1978}
W.R. {Ward} and A.G.W. {Cameron}, \emph{{Disc Evolution within the Roche
  Limit}}, in \emph{Lunar and Planetary Institute Conference Abstracts}, Mar.,
  1978, p. 1205.

\bibitem{Weaver2006}
H.A. {Weaver}, S.A. {Stern}, M.J. {Mutchler}, A.J. {Steffl}, M.W. {Buie}, W.J.
  {Merline}, J.R. {Spencer}, E.F. {Young}, and L.A. {Young}, Nature 439 (2006),
  pp. 943--945.

\bibitem{Showalter2011}
M.R. {Showalter}, D.P. {Hamilton}, S.A. {Stern}, H.A. {Weaver}, A.J. {Steffl},
  and L.A. {Young}, IAU Circ. 9221 (2011).

\bibitem{LeinhardtHaumea}
Z.M. Leinhardt, R.A. Marcus, and S.T. Stewart, The Astrophysical Journal 714
  (2010), p. 1789.

\bibitem{densities_paper}
A.C. Barr and M.E. Schwamb, Monthly Notices of the Royal Astronomical Society
  460 (2016), pp. 1542--1548,
  \urlprefix\url{http://dx.doi.org/10.1093/mnras/stw1052}.

\bibitem{Squyres88}
S.W. {Squyres}, R.T. {Reynolds}, A.L. {Summers}, and F.~{Shung}, J. Geophys.
  Res. 93 (1988), pp. 8779--8794.

\bibitem{Ward2010}
W.R. Ward and R.M. Canup, Astrophys. J. 140 (2010), pp. 1168--1193.

\bibitem{PapaloizouNelson2005}
J.C. {Papaloizou} and R.P. {Nelson}, Astron. and Astrophys. 433 (2005), pp.
  247--265.

\bibitem{Hubickyj2005}
O.~Hubickyj, P.~Bodenheimer, and J.J. Lissauer, Icarus 179 (2005), pp.
  415--431.

\bibitem{MolliereMordasini2012}
P.~Molli\`{e}re and C.~Mordasini, Astron. Astrophys. 547 (2012), p. A105.

\bibitem{Mordasini2013}
C.~Mordasini, Astron. Astrophys. 558 (2013), p. A113.

\bibitem{Haisch2001}
K.E. {Haisch} Jr., E.A. {Lada}, and C.J. {Lada}, {Astrophys. J. Lett.} 553
  (2001), pp. L153--L156.

\bibitem{Thi}
W.F. {Thi}, G.A. {Blake}, E.F. {van Dishoeck}, G.J. {van Zandelhoff}, J.M.M.
  {Horn}, E.E. {Becklin}, V.~{Mannings}, A.I. {Sargent}, M.E. {van den Ancker},
  and A.~{Natta}, Nature 409 (2001), pp. 60--63.

\bibitem{Lubow99}
S.H. {Lubow}, M.~{Seibert}, and P.~{Artymowicz}, Astrophys. J. 526 (1999), pp.
  1001--1012.

\bibitem{DAngelo2003}
G.~{D'Angelo}, T.~{Henning}, and W.~{Kley}, Astrophys. J. 599 (2003), pp.
  548--576.

\bibitem{AyliffeBatePaper2}
B.A. Ayliffe and M.R. Bate, Monthly Notices of the Royal Astronomical Society
  397 (2009), pp. 657--665.

\bibitem{TanigawaPaper1}
T.~Tanigawa, K.~Ohtsuki, and M.N. Machida, Astrophys. J. 747 (2012), p.~47.

\bibitem{TanigawaPaper2}
T.~Tanigawa, A.~Maruta, and M.N. Machida, Astrophys. J. 784 (2014), p. 109.

\bibitem{PapaloizouLarwood}
J.~Papaloizou and J.~Larwood, Monthly Notices of the Royal Astronomical Society
  315 (2000), pp. 823--833.

\bibitem{TanakaWard2004}
H.~Tanaka and W.R. Ward, The Astrophysical Journal 602 (2004), p. 388.

\bibitem{CresswellNelson}
P.~Cresswell and R.P. Nelson, Astron. Astrophys. 482 (2008), pp. 677--690.

\bibitem{PS00}
M.E. {Pritchard} and D.J. {Stevenson}, \emph{{Thermal Aspects of a Lunar Origin
  by Giant Impact}}, Origin of the earth and moon, edited by R.M.~Canup and
  K.~Righter and 69 collaborating authors.~Tucson: University of Arizona
  Press., p.179-196 (2000), pp. 179--196.

\bibitem{Cuk2016}
M.~{\'C}uk, L.~Dones, and D.~Nesvorn{\`y}, The Astrophysical Journal 820
  (2016), p.~97.

\bibitem{LukeRings}
L.~Dones, Icarus 92 (1991), pp. 194--203.

\bibitem{Robbins2010}
S.J. Robbins, G.R. Stewart, M.C. Lewis, J.E. Colwell, and
  M.~Srem{\v{c}}evi{\'c}, Icarus 206 (2010), pp. 431--445.

\bibitem{Cuzzi2010}
J.~Cuzzi, J.~Burns, S.~Charnoz, R.~Clark, J.~Colwell, L.~Dones, L.~Esposito,
  G.~Filacchione, R.~French, M.~Hedman, \emph{et~al.}, Science 327 (2010), pp.
  1470--1475.

\bibitem{CharnozLHB}
S.~Charnoz, A.~Morbidelli, L.~Dones, and J.~Salmon, Icarus 199 (2009), pp.
  413--428.

\bibitem{Canup2010}
R.M. Canup, Nature 468 (2010), pp. 943--946.

\bibitem{Charnoz2011}
S.~Charnoz, A.~Crida, J.C. Castillo-Rogetz, V.~Lainey, L.~Dones, O.~Karatekin,
  G.~Tobie, S.~Mathis, C.~Le~Poncin-Lafitte, and J.~Salmon, Icaurs 216 (2011),
  pp. 535--550.

\bibitem{RiederKenworthy2016}
S.~{Rieder} and M.A. {Kenworthy}, Astron. Astrophys. in press (2016),
  \urlprefix\url{http://arxiv.org/abs/1609.08485}.

\bibitem{Wood1986}
J.A. {Wood}, \emph{{Moon over Mauna Loa - A review of hypotheses of formation
  of Earth's Moon}}, in \emph{Origin of the Moon; Proceedings of the
  Conference, Kona, HI, October 13-16, 1984}, University of Arizona Press,
  Tucson, AZ, 1986, pp. 17--55.

\bibitem{Darwin1879}
G.H. Darwin, Philosophical Transactions of the Royal Society Part II 170
  (1879), pp. 447--530.

\bibitem{Jeffreys1930}
H.~{Jeffreys}, Monthly Notices of the Royal Astronomical Society 91 (1930), p.
  169.

\bibitem{Ringwood1960}
A.E. Ringwood, Geochimica et Cosmochimica Acta 20 (1960), pp. 241--259.

\bibitem{Wise1963}
D.U. Wise, Journal of Geophysical Research 68 (1963), pp. 1547--1554.

\bibitem{Darwin1880}
G.H. Darwin, Philosophical Transactions of the Royal Society of London 171
  (1880), pp. 713--891.

\bibitem{Stevenson1987}
D.J. Stevenson, Ann. Rev. Earth Planet Sci. 15 (1987), pp. 271--315.

\bibitem{DurisenScott1984}
R.H. Durisen and E.H. Scott, Icarus 58 (1984), pp. 153--158.

\bibitem{KaulaHarris1975}
W.M. Kaula and A.W. Harris, Reviews of Geophysics 13 (1975), pp. 363--371.

\bibitem{Nakazawa1983}
K.~Nakazawa, T.~Komuro, and C.~Hayashi, The moon and the planets 28 (1983), pp.
  311--327.

\bibitem{Pollack1979GasDrag}
J.B. Pollack, J.A. Burns, and M.E. Tauber, Icarus 37 (1979), pp. 587--611.

\bibitem{Opik1972}
E.J. {{\"O}pik}, Irish Astronomical Journal 10 (1972), p. 190.

\bibitem{WoodMitler}
J.A. {Wood} and H.E. {Mitler}, \emph{{Origin of the Moon by a Modified Capture
  Mechanism Or Half a Loaf is Better than a Whole One}}, in \emph{Lunar and
  Planetary Institute Conference Abstracts}, Mar., 1974, pp. 851--853.

\bibitem{MizunoBoss1985}
H.~Mizuno and A.~Boss, Icarus 63 (1985), pp. 109--133.

\bibitem{Tsui1999}
K.H. Tsui, Planetary and Space Science 47 (1999), pp. 917--920.

\bibitem{MurrayDermott}
C.D. {Murray} and S.F. {Dermott}, \emph{{Solar System Dynamics}}, Cambridge
  University Press, New York, 1999.

\bibitem{Meyer2010}
J.~Meyer, L.~Elkins-Tanton, and J.~Wisdom, Icarus 208 (2010), pp. 1--10.

\bibitem{Bennett2014}
D.P. {Bennett}, V.~{Batista}, I.A. {Bond}, C.S. {Bennett}, D.~{Suzuki}, J.P.
  {Beaulieu}, A.~{Udalski}, J.~{Donatowicz}, V.~{Bozza}, F.~{Abe}, C.S.
  {Botzler}, M.~{Freeman}, D.~{Fukunaga}, A.~{Fukui}, Y.~{Itow},
  N.~{Koshimoto}, C.H. {Ling}, K.~{Masuda}, Y.~{Matsubara}, Y.~{Muraki},
  S.~{Namba}, K.~{Ohnishi}, N.J. {Rattenbury}, T.~{Saito}, D.J. {Sullivan},
  T.~{Sumi}, W.L. {Sweatman}, P.J. {Tristram}, N.~{Tsurumi}, K.~{Wada}, P.C.M.
  {Yock}, {MOA Collaboration}, M.D. {Albrow}, E.~{Bachelet}, S.~{Brillant},
  J.A.R. {Caldwell}, A.~{Cassan}, A.A. {Cole}, E.~{Corrales}, C.~{Coutures},
  S.~{Dieters}, D.~{Dominis Prester}, P.~{Fouqu{\'e}}, J.~{Greenhill},
  K.~{Horne}, J.R. {Koo}, D.~{Kubas}, J.B. {Marquette}, R.~{Martin}, J.W.
  {Menzies}, K.C. {Sahu}, J.~{Wambsganss}, A.~{Williams}, M.~{Zub}, {PLANET
  Collaboration}, J.Y. {Choi}, D.L. {DePoy}, S.~{Dong}, B.S. {Gaudi},
  A.~{Gould}, C.~{Han}, C.B. {Henderson}, D.~{McGregor}, C.U. {Lee}, R.W.
  {Pogge}, I.G. {Shin}, J.C. {Yee}, {{$\mu$}FUN Collaboration}, M.K.
  {Szyma{\'n}ski}, J.~{Skowron}, R.~{Poleski}, S.~{Koz{\l}owski},
  {\L}.~{Wyrzykowski}, M.~{Kubiak}, P.~{Pietrukowicz}, G.~{Pietrzy{\'n}ski},
  I.~{Soszy{\'n}ski}, K.~{Ulaczyk}, {OGLE Collaboration}, Y.~{Tsapras}, R.A.
  {Street}, M.~{Dominik}, D.M. {Bramich}, P.~{Browne}, M.~{Hundertmark},
  N.~{Kains}, C.~{Snodgrass}, I.A. {Steele}, {RoboNet Collaboration},
  I.~{Dekany}, O.A. {Gonzalez}, D.~{Heyrovsk{\'y}}, R.~{Kandori}, E.~{Kerins},
  P.W. {Lucas}, D.~{Minniti}, T.~{Nagayama}, M.~{Rejkuba}, A.C. {Robin}, and
  R.~{Saito}, Astrophys. J. 785 (2014), 155.

\bibitem{HellerAbio2014}
R.~Heller, D.~Williams, D.~Kipping, M.A. Limbach, E.~Turner, R.~Greenberg,
  T.~Sasaki, E.~Bolmont, O.~Grasset, K.~Lewis, \emph{et~al.}, Astrobiology 14
  (2014), pp. 798--835.

\bibitem{WeidnerHorne2010}
C.~Weidner and K.~Horne, Astronomy \& Astrophysics 521 (2010), p. A76.

\bibitem{KippingKepler90}
D.M. Kipping, X.~Huang, D.~Nesvorn{\`y}, G.~Torres, L.A. Buchhave, G.{\'A}.
  Bakos, and A.R. Schmitt, The Astrophysical Journal Letters 799 (2015), p.
  L14.

\bibitem{KippingPOS}
D.M. {Kipping}, \emph{{In Search of Exomoons}}, in \emph{Frank N. Bash
  Symposium 2013: New Horizons in Astronomy, October 6-8, 2013, Austin, TX},
  Oct., \urlprefix\url{http://arxiv.org/abs/1405.1455}, 2014.

\bibitem{Agol2015}
E.~{Agol}, T.~{Jansen}, B.~{Lacy}, T.D. {Robinson}, and V.~{Meadows},
  Astrophys. J. 812 (2015), p.~5.

\bibitem{Boss97}
A.P. {Boss}, Science 276 (1997), pp. 1836--1839.

\bibitem{Williams2013}
D.M. {Williams}, Astrobiology 13 (2013), pp. 315--323.

\bibitem{PorterGrundy2011}
S.B. {Porter} and W.M. {Grundy}, Astrophys. J. 736 (2011), p. L14.

\bibitem{MacDonald1966}
G.J.F. {MacDonald}, \emph{{Origin of the Moon: Dynamical Considerations}}, in
  \emph{The Earth-Moon System: Proceedings of an International Conference,
  January 20-21, 1964}, Plenum Press, New York, 1966, pp. 165--210.

\bibitem{Brozovic}
M.~Brozovi{\'c}, M.R. Showalter, R.A. Jacobson, and M.W. Buie, Icarus 246
  (2015), pp. 317--329.

\end{thebibliography}
\end{document}